\documentclass[10pt,a4paper]{article}
\usepackage[a4paper, left=2cm, right=2cm, top=2cm]{geometry}

\usepackage[utf8]{inputenc}
\usepackage{amsmath, amsthm, amssymb, amsfonts}     
\usepackage{mathrsfs}                               
\usepackage{accents}                                
\usepackage{natbib}                                 
\usepackage{hyperref}                               
\usepackage{doi}                                    
\usepackage{authblk}                                
\usepackage{siunitx}                                
\usepackage{booktabs}                               
\usepackage{ulem}                                   
\usepackage{nomencl}                                
\makenomenclature                                   
\usepackage[linesnumbered,ruled,vlined]{algorithm2e}
\SetCommentSty{mycommfont}
\usepackage{subfigure}

\usepackage{bm}                     

\usepackage{graphicx}
\usepackage{tikz}
\usepackage{pgfplots}

\usetikzlibrary{patterns}

\usetikzlibrary{decorations.pathmorphing,arrows,intersections,arrows.meta,angles,quotes}

\pgfplotsset{
	kurze Legende/.style={
		legend image code/.code={
			\draw[##1,mark repeat=2,line width=0.6pt]
			plot coordinates {
				(0cm,0cm)
				(0.3cm,0cm)
			};
		}
	}
}

\pgfplotsset{
	compat = newest,
	scale only axis, 
	max space between ticks = 50pt,
	ticklabel style = {font=\footnotesize},
	legend style =  {font=\footnotesize},
	grid = major,
	grid style = {dotted},
	legend columns=1, 
	xtick pos=left,
	ytick pos=left
}

\pgfplotsset{select coords between index/.style 2 args={
		x filter/.code={
			\ifnum\coordindex<#1\fi
			\ifnum\coordindex>#2\fi
		}
}}

\definecolor{color1}{HTML}{0060AD} 
\definecolor{color2}{HTML}{FF4500} 
\definecolor{color3}{HTML}{FFA500} 
\definecolor{color4}{HTML}{006400} 
\definecolor{color5}{HTML}{9400D3} 
\definecolor{color6}{HTML}{800000} 
\definecolor{color7}{HTML}{000000} 
\definecolor{color8}{HTML}{0000FF} 
\definecolor{color9}{HTML}{FF0000} 
\definecolor{mycolor_blue}{RGB}{66,124,161}
\definecolor{mycolor_grey}{RGB}{198,198,198} 

\tikzstyle{line1} = [color=color7,semithick] 
\tikzstyle{line2} = [color=color2,densely dotted,semithick]
\tikzstyle{line3} = [color=color1,densely dashed,semithick]
\tikzstyle{line4} = [color=color5,dash dot,semithick]
\tikzstyle{line5} = [color=color4,dash dot dot,semithick]
\tikzstyle{line6} = [color=color6,semithick]
\tikzstyle{line7} = [color=color8,densely dotted,semithick]

\tikzstyle{mark1} = [color=color7,mark=x,mark size=2pt,mark options=solid,semithick] 
\tikzstyle{mark2} = [color=color2,mark=o,mark size=2pt,mark options=solid,semithick]
\tikzstyle{mark3} = [color=color1,mark=*,mark size=2pt,mark options=solid,semithick]
\tikzstyle{mark4} = [color=color5,mark=triangle,mark size=2pt,mark options=solid,semithick]
\tikzstyle{mark5} = [color=color4,mark=square,mark size=2pt,mark options=solid,semithick]
\tikzstyle{mark6} = [color=color7,mark=o,mark size=2pt,mark options=solid,semithick]
\tikzstyle{mark7} = [color=color7,mark=*,mark size=2pt,mark options=solid,semithick]
\tikzstyle{mark8} = [color=color7,mark=triangle,mark size=2pt,mark options=solid,semithick]

\usetikzlibrary{external}
\title{Elliptic Relaxation Strategies to Support Numerical Stability of Segregated Continuous Adjoint Flow Solvers}
\author[1]{Niklas K\"uhl\thanks{niklas.kuehl@tuhh.de}}

\affil[1]{Hamburg Ship Model Basin, Bramfelder Strasse 164, D-22305 Hamburg, Germany}

\begin{document}

\providetoggle{tikzExternal}
\settoggle{tikzExternal}{true}
\settoggle{tikzExternal}{false}

\maketitle

\begin{abstract}
This paper introduces a novel method for numerically stabilizing sequential continuous adjoint flow solvers utilizing an elliptic relaxation strategy. The proposed approach is formulated as a Partial Differential Equation (PDE) containing a single user-defined parameter, which analytical investigations reveal to represent the filter width of a probabilistic density function or Gaussian kernel. Key properties of the approach include (a) smoothing features with redistribution capabilities while (b) preserving integral properties. The technique targets explicit adjoint cross-coupling terms, such as the Adjoint Transpose Convection (ATC) term, which frequently causes numerical instabilities, especially on unstructured grids common in industrial applications. A trade-off is made by sacrificing sensitivity consistency to achieve enhanced numerical robustness.

The method is validated on a two-phase, laminar, two-dimensional cylinder flow test case at $\mathrm{Re}_\mathrm{D} = 20$ and $\mathrm{Fn} = 0.75$, focusing on minimizing resistance or maximizing lift. A range of homogeneous and inhomogeneous filter widths is evaluated. Subsequently, the relaxation method is employed to stabilize adjoint simulations during shape optimizations that aim at drag reduction of ship hulls. Two case studies are considered: A model-scale bulk carrier traveling at $\mathrm{Re}_\mathrm{L} = \SI{7.246}{} \cdot 10^6$ and $\mathrm{Fn}=0.142$ as well as a harbor ferry cruising at $\mathrm{Re}_\mathrm{L} = \SI{2.43}{} \cdot 10^8$ and $\mathrm{Fn}=0.4$ in full-scale conditions. Both cases, characterized by unstructured grids prone to adjoint divergence, demonstrate the effectiveness of the proposed method in overcoming stability challenges. The resulting optimizations achieve superior outcomes compared to approaches that omit problematic coupling terms.
\end{abstract}

\begin{flushleft}
\end{flushleft}

\section{Introduction}

Computational engineering, now and in the future, should do more than analyze engineering designs; it should also strive to optimize them. Due to their modularity, Computational Fluid Dynamics (CFD) techniques with sequential or generally fixed-point iteration-based solution strategies feature comparatively high user flexibility and have established themselves as robust prediction tools in various disciplines. The incorporation of desirable adjoint methods for optimization purposes, therefore, has the fate of adapting to the sequential primal simulation process, either in line with the discrete (\cite{nadarajah2003discrete, nielsen2004implicit, roth2013discrete} or the continuous adjoint method (\cite{nadarajah2000comparison, othmer2008continuous, peter2010numerical}), whereby, for consistency reasons, as many discrete building blocks of the primal solution process as possible should be used, partly modified by situational, e.g., sign switching or mirroring of interpolation schemes (\cite{nadarajah2000comparison, stuck2013adjoint}). Due to the linearized approach, adjoint investigations of highly non-linear, e.g., multiphase turbulent flows, feature --in addition to the primal term count-- several further adjoint contributions, that couple the adjoint equations among themselves, cf. \cite{kuhl2022discrete, kuhl2022adjoint}. However, these additional terms must often be treated explicitly in the underlying sequential solution process. Consequently, the linear adjoint system of equations might feature a severe source term dominance and thus numerical stiffness, which might impede the solution approximation that is not necessarily better conditioned than that of the non-linear primal system. A classic example refers to the debate on the "if and how" consideration of the Adjoint Transpose Convection (ATC) term, cf. \cite{lohner2003adjoint, othmer2014adjoint, karpouzas2016adjoint}. Additionally, the reversed information transport of dual transport processes may favor different grid point distributions. This compromise of numerical robustness can already be challenging when considering academic problems but makes adjoint methods in industrial problems much more difficult, usually further exacerbated by the utilization of unstructured spatial discretizations and, in this respect, influenced, partly erratic gradient representations. Various strategies for dealing with the numerical robustness mentioned above are presented in the literature, e.g., via (a) hard consistency breaks by neglecting all or selected, numerically particularly unpleasant source terms (\cite{lohner2003adjoint, othmer2008continuous, kuhl2021continuous}), (b) strong under relaxation, for example, on the basis of an integration of the steady adjoint equations in a pseudo-time (\cite{kroger2018adjoint, kuhl2021adjoint}), or (c) the introduction of additional stability-enhancing terms, e.g. by introducing a solid amount of apparent viscosity (\cite{beckers2019duality, kuhl2021adjoint}).

This manuscript presents an alternative method for promoting numerical stability during sequential continuous adjoint CFD studies. The strategy is based on a local redistribution process aligned with an elliptic relaxation procedure, which inheres smoothing capabilities while featuring integral conservation of considered adjoint source terms. The approach has been used for the regularization of iterative solution processes (\cite{grossmann1997smoothing}) and inverse problems (\cite{al2018smoothing}) or to enhance flow turbulence modeling (\cite{durbin1991near}). The strategy also received attention in topology optimization, cf. \cite{lazarov2011filters, lazarov2016length, galanos2022synergistic}. In the field of CFD-based shape optimization, such approaches usually gather under the umbrella of "Sobolev smoothing," cf. \cite{jameson2003aerodynamic, kim2005enhancement, mohammadi2010applied, kroger2015cad, dick2022combining}. The implicit approach is modeled via a Partial Differential Equation (PDE) equipped with a single, user-adjustable parameter, which can be interpreted as an inhomogeneous squared length measure that allows control of the relaxation, for example, by coupling to global reference lengths or local grid dimensions. It is to be expected that large filter widths tend to reduce the predictive quality of the overall adjoint solution process. The trade-off between (a) increased numerical robustness and (b) reduced dual consistency due to the elliptic relaxation of adjoint-coupling terms is deliberately accepted. This paper investigates the proposed method on adjoint two-phase frozen-turbulence Reynolds-Averaged Navier-Stokes (RANS) flows, which already exhibit several numerically unpleasant coupling terms. However, due to the basic PDE modeling, the presented idea can be adapted to many PDE-solving computational software packages and thus applied to similar, e.g., coupled problems.

The manuscript is structured as follows. The upcoming Sec. \ref{sec:mathematical_model} considers the required primal and adjoint balance equations, emphasizing source term dominance and equation coupling aspects. Subsequently, Sec. \ref{sec:elliptic_relaxation} introduces the relaxation equation, including boundary conditions, and discusses the fundamental mathematical properties of the approach. Section \ref{sec:validation} verifies and validates the elliptic relaxation method using a laminar two-phase flow around a circular cylinder at $\mathrm{Re}_\mathrm{D} = 20$ and $\mathrm{Fn} = 0.75$. Section \ref{sec:application} applies the stabilization method to complex three-dimensional optimizations of the hulls of a bulk carrier traveling at $\mathrm{Re}_\mathrm{L} = \SI{7.246}{} \cdot 10^6$ and $\mathrm{Fn}=0.142$ as well as a harbor ferry cruising at $\mathrm{Re}_\mathrm{L} = \SI{2.43}{} \cdot 10^8$ and $\mathrm{Fn}=0.4$. The paper concludes with a summary in Sec. \ref{sec:conclusion}. The manuscript defines vectors and tensors based on a Cartesian coordinate system, with Einstein's summation convention applied to doubly occurring subscripts. 


\section{Primal and Adjoint Flow Model}
\label{sec:mathematical_model}

The paper's governing fluid dynamic model refers to a classical, statistically averaged flow concept based on the RANS approach applied to an incompressible viscous ($\mu$) fluid with density $\rho$, closed with a two-equation Boussinesq-viscosity model and coupled to a compressive Volume-of-Fluid (VoF) formulation. The following balance equations describe the evolution of the fluid velocity vector $v_\mathrm{i}$, its pressure $p$, and volume fraction $c$ in residual $r^\mathrm{(\cdot)}$ form, viz.
\begin{alignat}{2}
    r_\mathrm{i}^\mathrm{v} &= 0 &&= \frac{\partial \rho v_\mathrm{i}}{\partial t} + \frac{\partial v_\mathrm{k} \, \rho \, v_\mathrm{i}}{\partial x_\mathrm{k}}  + \frac{\partial }{\partial x_\mathrm{k}} \bigg[ p^\mathrm{eff} \delta_\mathrm{ik} - \mu^\mathrm{eff} 2 S_\mathrm{ik} \bigg] - \rho g_\mathrm{i} \label{equ:primal_momentum_balance} \\
    r^\mathrm{p} &= 0 &&= -\frac{\partial v_\mathrm{k}}{\partial x_\mathrm{k}} \label{equ:primal_mass_balance} \\
    r^\mathrm{c} &= 0 &&= \frac{\partial c}{\partial t}+ \frac{\partial v_\mathrm{k} \, c}{\partial x_\mathrm{k}} \label{equ:primal_vof_balance} \, .
\end{alignat}
The unit coordinates, and the strain rate tensor are denoted by the Kronecker Delta $\delta_\mathrm{ik}$ and $S_\mathrm{ik} = 1/2 (\nabla_\mathrm{k} v_\mathrm{} + \nabla_\mathrm{i} v_\mathrm{k})$.
In \eqref{equ:primal_momentum_balance}, $\mu^\mathrm{eff} = \mu + \mu^\mathrm{t}$ as well as $p^\mathrm{eff} = p + p^\mathrm{t}$ represent effective viscosity and pressure of a fully turbulent, statistically averaged ($\mu^\mathrm{t} \to \rho C_\mu k^2 / \varepsilon$, $p^\mathrm{t} \to (2/3) \rho k$) fluid flow, respectively. Details on the computation of $p^\mathrm{t}$ and $\mu^\mathrm{t}$ refer to standard two-equation closure practices and are omitted to save space. Further information can be found, e.g., in \cite{pope2001turbulent, wilcox1998turbulence} regarding theoretical background or \cite{stuck2012adjoint, manzke2018development, kuhl2021phd} for practical implementation aspects. Laminar flow studies skip the turbulent modeling part and neglect $p^\mathrm{t} \to 0$ and $\mu^\mathrm{t} \to 0$.
Local fluid properties follow the volume fraction field based on an interpolation between the fluid's bulk properties ($\rho_a$, $\mu_a$) and ($\rho_b$, $\mu_b$), i.e.,
\begin{align}
    \rho = \rho_a c + (1-c) \rho_b
    \qquad \qquad
    \mu = \mu_a c + (1-c) \mu_b \, .
\end{align}

Adjoint procedures are usually derived based on a cost functional to be optimized. A general integral objective functional $J$ which consists of boundary ($j^\mathrm{\Gamma}$) and volume ($j^\mathrm{\Omega}$) contributions along the objective surface $\Gamma^\mathrm{O}$ and the objective volume $\Omega^\mathrm{O}$ can be defined as follows
\begin{align}
    J = \int \int_\mathrm{\Gamma^\mathrm{O}} j^\mathrm{\Gamma} \mathrm{d} \Gamma + \int_\mathrm{\Omega^\mathrm{O}} j^\mathrm{\Omega} \mathrm{d} \Omega \mathrm{d} t \label{equ:general_objective}
    \qquad \qquad \to \qquad \qquad
    L = J  + \int \int_\Omega ( \hat{v}_\mathrm{i} \, r_\mathrm{i}^\mathrm{v_\mathrm{i}} + \hat{p} \, r^\mathrm{p} + \hat{c} \, r^\mathrm{c}) \, \mathrm{d} \Omega \mathrm{d} t \, .
\end{align}
The objective has already been supplemented by the primal residuals (\ref{equ:primal_momentum_balance}), (\ref{equ:primal_mass_balance}) and (\ref{equ:primal_vof_balance}) to form an augmented objective functional, often referred to as a Lagrange functional. The introduced Lagrangian multipliers refer to the adjoint velocity $\hat{v}_\mathrm{i}$, the adjoint pressure $\hat{p}$, and the adjoint volume fraction quantity $\hat{c}$. Adjoint turbulence variables are deliberately neglected in line with the frozen turbulence assumption, cf. \cite{othmer2008continuous, stuck2013adjoint, marta2013handling}. Differentiation of the Lagrangian in the direction of the primal state followed by the application of first-order optimality conditions provide the adjoint equations to \eqref{equ:primal_momentum_balance}-\eqref{equ:primal_vof_balance}. An interested reader in the derivation might consult, e.g., \cite{lohner2003adjoint, othmer2008continuous, stuck2013adjoint, kroger2018adjoint, kuhl2021adjoint}. The adjoint equations are directly notated to save space, viz.
\begin{alignat}{2}
    r_\mathrm{i}^\mathrm{\hat{v}} &= 0 &&=  -\frac{\partial \rho \hat{v}_\mathrm{i}}{\partial t}  - \frac{\partial  v_\mathrm{k} \,  \rho \, \hat{v}_\mathrm{i}}{\partial x_\mathrm{k}} + \frac{\partial }{\partial x_\mathrm{k}} \bigg[ \hat{p} \delta_\mathrm{ik} - \mu^\mathrm{eff} 2 \hat{S}_\mathrm{ik} \bigg] + \hat{v}_\mathrm{k} \rho \frac{\partial v_\mathrm{k}}{\partial x_\mathrm{i}} + \hat{c} \frac{\partial c}{\partial x_\mathrm{i}} + \frac{\partial j^\mathrm{\Omega}}{\partial v_\mathrm{i}} \label{equ:adjoint_momentum_balance} \, , \\
    r^\mathrm{\hat{p}} &= 0 &&= -\frac{\partial \hat{v}_\mathrm{k}}{\partial x_\mathrm{k}} + \frac{\partial j^\mathrm{\Omega}}{\partial p} \label{equ:adjoint_mass_balance} \, , \\
    r^\mathrm{\hat{c}} &= 0 &&= \frac{\partial \hat{c}}{\partial t} - \frac{\partial v_\mathrm{k} \, \hat{c}}{\partial x_\mathrm{k}} + \rho_\Delta \left[ \hat{v}_\mathrm{i} v_\mathrm{k} \frac{\partial v_\mathrm{i}}{\partial x_\mathrm{k}} - \hat{v}_\mathrm{i} g_\mathrm{i} + \frac{\partial \mu^\mathrm{t}}{\partial \rho} 2 S_\mathrm{ik} \hat{S}_\mathrm{ik} \right] + \mu_\Delta 2 S_\mathrm{ik} \frac{\partial \hat{v}_\mathrm{i}}{\partial x_\mathrm{k}} + \frac{\partial j^\mathrm{\Omega}}{\partial c} \label{equ:adjoint_vof_balance} \, .
\end{alignat}
The adjoint strain rate tensor $\hat{S}_\mathrm{ik} = 1/2 (\nabla_\mathrm{k} \hat{v}_\mathrm{i} + \nabla_\mathrm{i} \hat{v}_\mathrm{k})$ has been introduced.
The adjoint Eqns. \eqref{equ:adjoint_momentum_balance}-\eqref{equ:adjoint_vof_balance} feature several additional source terms compared to their primal complements in Eqns. \eqref{equ:primal_momentum_balance}-\eqref{equ:primal_vof_balance}. These sources might impede the sequential adjoint solution process, especially the adjoint pressure-velocity coupling, and the upcoming Sec. \ref{sec:elliptic_relaxation} presents a relaxation strategy to improve the numerical robustness of the adjoint solution approximation process.

After successfully approximating the primal and adjoint equations, remaining optimality requirements allow the determination of sensitivity rules with respect to general control parameters. The habitat of these sensitivities depends on the underlying optimization problem but can roughly be generalized to have topological (\cite{borrvall2003topology, othmer2008continuous}) and geometric (\cite{giles2000introduction, stuck2013adjoint}) characteristics. In the context of this paper, the latter is examined, i.e., certain walled boundaries of the flow domain are released for design so that the following sensitivity rule arises
\begin{align}
    \delta_\mathrm{u} J = \int_\mathrm{\Gamma^\mathrm{D}} s \mathrm{d} \Gamma
    \qquad \qquad \mathrm{with} \qquad \qquad
    s = - \mu \frac{\partial v_\mathrm{i}}{\partial n} \frac{\partial \hat{v}_\mathrm{i}}{\partial n} \, . \label{equ:sensitivity_derivative}
\end{align}

\paragraph{The Utilized Segregated Numerical Procedure} to approximate the PDEs of the primal and adjoint systems, as well as the later proposed elliptic relaxation procedure, is built upon the Finite-Volume method FreSCo+ \cite{rung2009challenges, stuck2012adjoint, kroger2016numerical, kuhl2021phd}. Similar to integration-by-parts in deriving the continuous adjoint equations, summation-by-parts is applied to derive the components of the discrete adjoint expressions. A comprehensive derivation of this hybrid adjoint approach can be found in \cite{stuck2013adjoint, kroger2016numerical} for the single-phase system and in \cite{kroger2018adjoint, kuhl2021adjoint, kuhl2022adjoint} for the multiphase system. The implicit numerical scheme is second-order accurate and allows the use of general polyhedral cells. Convective primal [adjoint] momentum fluxes are approximated using the Quadratic Upstream [Downstream] Interpolation for Convective Kinematics QUICK [QDICK] scheme, cf. \cite{stuck2013adjoint}. The primal and adjoint pressure-velocity coupling is based on the Semi Implicit Method for Pressure-Linked Equations (SIMPLE), with parallelization achieved through domain decomposition. For parameter-free shape optimization, the computational grid can be modified using a Laplace-Beltrami \cite{kroger2015cad} or Steklov-Poincaré \cite{haubner2020analysis, kuhl2022adjoint} type surface metric approach.
The discrete approximation of primal and adjoint pressure gradients follows a thoroughly conservative approach based on integration by parts via Gauss's Theorem. Remaining gradient approximations utilize a non-conservative but more robust weighted Least-Squares method.
A compressive convective concentration approximation scheme keeps the fluid interface sharp. As a result, even for steady applications like, e.g., the calm water resistance of a vessel, the solution advances in a pseudo-time. It is terminated once temporal changes are negligible, e.g., $\partial c / \partial t \to 0$ in Eqn. \eqref{equ:primal_vof_balance}. Consequently, the cost functional and the Lagrangian include the physical time, but typical issues of transient adjoint approaches like memory overheads are not relevant \cite{margetis2021lossy}. Further information on interpreting the primal/adjoint time horizon of free-surface flows is provided in \cite{kuhl2021adjoint, kuhl2022adjoint}. However, the suggested stabilization strategy could also be applied to unsteady adjoint-based optimization problems.

\section{Numerical Stabilization Model}
\label{sec:elliptic_relaxation}

The additional, sometimes with significant fluid property variations, scaled additional terms on the right-hand side of the adjoint equations ensure a strong source term dominance and, therefore, a stiff equation system for the underlying sequential adjoint solution process. For instance, the two-phase contribution to the adjoint momentum balance ($\hat{c} \nabla_\mathrm{i} c$) must inevitably be considered an explicit source. The same applies to large parts of the ATC term ($\rho \hat{v}_\mathrm{k} \nabla_\mathrm{i} v_k$). Explicit adjoint source terms usually contain gradient expressions, which can exhibit physically (e.g., by the phase jump along the water-air interface) and numerically (e.g., due to utilized unstructured grids) motivated volatility. The resulting unpleasant explicit contributions negatively affect the numerical behavior of the sequential adjoint solver.

The following linear, elliptic, scalar-valued PDE provides the basis for the paper's stability-enhancing treatment of adjoint source terms. A generic right-hand side $\varphi$ represents the source or input for a relaxed companion or output field $\tilde{\varphi}$, viz.
\begin{alignat}{2}
    \tilde{\varphi} - \frac{\partial}{\partial x_\mathrm{k}} \left[ \mu_\varphi \frac{\partial \tilde{\varphi}}{\partial x_\mathrm{k}} \right] &= \varphi \qquad &&\text{in} \qquad \Omega \label{equ:elliptic_relaxation} \\
     \mu_\varphi \frac{\partial \tilde{\varphi}}{\partial n} &= 0 \qquad &&\text{on} \qquad \Gamma
\end{alignat}
where $\tilde{\varphi}$ corresponds to a smoothed version of the r.h.s. $\varphi$ based on a positive and for dimensional reasons squared filter width $[\mu_\varphi] = \SI{}{m^2}$. The diffusivity can be chosen either homogeneously (e.g., linked to a global reference length) or in-homogeneously (e.g., scaled by the local grid spacing). The relaxation strategy can be applied to various primal and adjoint expressions and terms. However, this article only applies the relaxation to sources of adjoint equations. Additionally, skillful manipulation and specification of different, e.g., zero Dirichlet boundary conditions might favor zero solutions in desired areas -- this will not be considered in this paper.
Equation \eqref{equ:elliptic_relaxation} features smoothing and filtering characteristics that fade out $\tilde{\varphi} \to \varphi$ for small filter widths and vanish in the limiting case of $\mu_\varphi \to 0$.

\begin{figure}[!ht]
\centering
\subfigure[]{
\iftoggle{tikzExternal}{
\input{./tikz/finite_volume_approximation_field.tikz}}{
\includegraphics{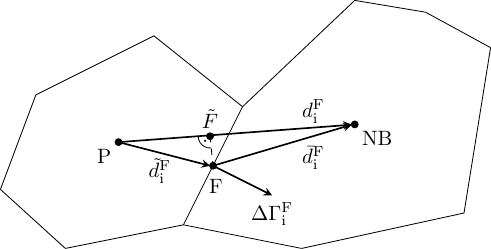}}
}
\hspace{2cm}
\subfigure[]{
\iftoggle{tikzExternal}{
\input{./tikz/finite_volume_approximation_boundary.tikz}}{
\includegraphics{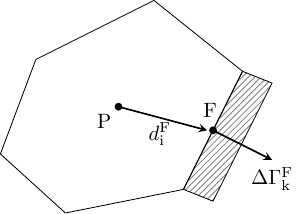}}
}
\caption{Schematic representation of a Finite-Volume arrangement (a) in the field and (b) along the boundary.}
\label{fig:finite_volume_approximation}
\end{figure}

The linear Eqn. \eqref{equ:elliptic_relaxation} can be approximated by standard means of modern CFD packages (assembling, solving, etc.), ideally aiming at a fast solution process with minimal iteration effort due to, e.g., under-relaxation. This paper's face and thus flux-based Finite-Volume-Method (FVM) assembles the necessary Laplace operator to treat potential non-orthogonality contributions explicitly via the right-hand side, cf. \cite{stuck2012adjoint, kuhl2021phd}. These contributions are deliberately neglected for convergence acceleration and computational overhead minimization. Further information on the employed FVM code FreSCo$^+$ can be found, e.g., in \cite{rung2009challenges, stuck2012adjoint, manzke2018development, schubert2019analysis, kuhl2021adjoint}.
Variable, inhomogeneous filter widths that scale with the local grid spacing assign the length scale via the distance vector of a face F connecting the adjacent cell's P and NB centroids, cf. Fig. \ref{fig:finite_volume_approximation}. The generic implementation reads
\begin{align}
    \mu_\varphi^\mathrm{F} = A \cdot |d_\mathrm{i}^\mathrm{F}|^B \, .
\end{align}
While the exponent is consistently assigned to $B = 2$, the scaling variable $A$ is varied throughout the paper. The discrete length scale is denoted as $|d_\mathrm{i}^\mathrm{F}| \to \Delta x$ to shorten the notation in the remainder of the paper.

\paragraph{Equation Properties} are discussed in the following. For a homogeneous viscosity, the following 1D solution of Eqn. \eqref{equ:elliptic_relaxation} in $x_\mathrm{1}$ direction can be constructed, viz.
\begin{align}
    \tilde{\varphi} &= e^{-\frac{x}{\sqrt{\mu_\varphi}}} \int_1^x \frac{e^{\frac{\xi}{\sqrt{\mu_\varphi}}}}{2 \sqrt{\mu_\varphi}} \varphi(\xi) \mathrm{d} \xi - e^{\frac{x}{\sqrt{\mu_\varphi}}} \int_1^x \frac{e^{-\frac{\xi}{\sqrt{\mu_\varphi}}}}{2 \sqrt{\mu_\varphi}} \varphi(\xi) \mathrm{d} \xi + C_1 e^{\frac{x_\mathrm{1}}{\sqrt{\mu_\varphi}}} + C_2 e^{-\frac{x_\mathrm{1}}{\sqrt{\mu_\varphi}}} \, . \label{eqn:er_1d_sol}
\end{align}
The generalization to the third dimension supplies the following solution
\begin{align}
    \tilde{\varphi} &= e^{-\frac{{x_1}}{\sqrt{\mu_\varphi}}} \int_1^{x_1} \frac{e^{\frac{\xi}{\sqrt{\mu_\varphi}}}}{6 \sqrt{\mu_\varphi}} \varphi(\xi) \mathrm{d} \xi - e^{\frac{{x_1}}{\sqrt{\mu_\varphi}}} \int_1^{x_1} \frac{e^{-\frac{\xi}{\sqrt{\mu_\varphi}}}}{6 \sqrt{\mu_\varphi}} \varphi(\xi) \mathrm{d} \xi \\
    &+ e^{-\frac{{x_2}}{\sqrt{\mu_\varphi}}} \int_1^{x_2} \frac{e^{\frac{\xi}{\sqrt{\mu_\varphi}}}}{6 \sqrt{\mu_\varphi}} \varphi(\xi) \mathrm{d} \xi - e^{\frac{{x_2}}{\sqrt{\mu_\varphi}}} \int_1^{x_2} \frac{e^{-\frac{\xi}{\sqrt{\mu_\varphi}}}}{6 \sqrt{\mu_\varphi}} \varphi(\xi) \mathrm{d} \xi \\
    &+ e^{-\frac{{x_3}}{\sqrt{\mu_\varphi}}} \int_1^{x_3} \frac{e^{\frac{\xi}{\sqrt{\mu_\varphi}}}}{6 \sqrt{\mu_\varphi}} \varphi(\xi) \mathrm{d} \xi - e^{\frac{{x_3}}{\sqrt{\mu_\varphi}}} \int_1^{x_3} \frac{e^{-\frac{\xi}{\sqrt{\mu_\varphi}}}}{6 \sqrt{\mu_\varphi}} \varphi(\xi) \mathrm{d} \xi \\
    & + C_1 e^{\frac{{x_1} + {x_2} + {x_3}}{\sqrt{\mu_\varphi}}} + C_2 e^{-\frac{{x_1} + {x_2} + {x_3}}{\sqrt{\mu_\varphi}}} \, .
\end{align}
The solutions are composed of bell-function-like expressions that indicate convolutional characteristics. Later numerical results show Eqn. \eqref{equ:elliptic_relaxation}'s almost perfect conservative character even for inhomogeneous viscosities, which is only conjecturable based on Eqn. \eqref{eqn:er_1d_sol} but is emphasized by the following rational, inspired by \cite{stuck2011adjoint, kroger2015cad}. Assuming that the relaxation input on the r.h.s. of Eqn. \eqref{equ:elliptic_relaxation} can be interpreted as a pointed localized, instantaneous source, the initial equation can be rewritten as follows by assuming a (pseudo) temporal dependence of the variables (e.g., $\varphi(x_i) \to \varphi(x_i,t)$) accompanied with a homogeneous but transient viscosity $\mu_\varphi \to t h^2 / 16$, viz.
\begin{align}
    \tilde{\varphi} - \frac{\partial}{\partial x_\mathrm{k}} \left[ \mu \frac{\partial \tilde{\varphi}}{\partial x_\mathrm{k}} \right] = \varphi
    \qquad \qquad \to \qquad \qquad
    \tilde{\varphi}(t) - \frac{\partial}{\partial x_\mathrm{k}} \left[ \frac{t h^2}{16} \frac{\partial \tilde{\varphi}(t)}{\partial x_\mathrm{k}} \right] = \varphi(t=0) + \mathcal{O}(t) \, ,
\end{align}
whereby a temporally constant first-order approximation has been introduced for the source term which disappears when computing a temporal derivative. In doing so, a homogeneous heat equation PDE is obtained. The following exponential function gives its solution
\begin{align}
    \frac{\partial \tilde{\varphi}(t)}{\partial t} - \frac{\partial}{\partial x_\mathrm{k}} \left[ \frac{h^2}{16} \frac{\partial \tilde{\varphi}(t)}{\partial x_\mathrm{k}} \right] = 0
    \qquad \qquad \to \qquad \qquad
     \tilde{\varphi}(t) &= \frac{1}{( 0.5 \sqrt{\pi} h)^D \sqrt{t}} e^{-\frac{4}{t} \left( \frac{x_\mathrm{k} - \tilde{x}_\mathrm{k}}{h} \right)^2} \, , \label{equ:gaussian}
\end{align}
frequently labeled as heat kernel. The parameters $1 \leq D \leq 3$ and $h = 2 \sqrt{2} \sigma$ denote the spatial dimension and a measure of the kernel width or filter length that is related to the standard deviation $\sigma$ of a probability density function.
The solution in Eqn. \eqref{equ:gaussian} resembles a Gaussian kernel function that features, e.g., compact support ($\tilde{\varphi}(t,x_i,\tilde{x}_\mathrm{i} \to \infty) \to 0$), symmetry ($\tilde{\varphi}(t, x_i, \tilde{x}_i) = \tilde{\varphi}(t, x_i, -\tilde{x}_i)$), and normalization ($\int \tilde{\varphi}(t,x_i,\tilde{x}_i) \mathrm{d} \tilde{x}_\mathrm{i} = 1$) properties. The latter underlines the conservative character of the proposed elliptic relaxation approach. As a consequence, an alternative to implicitly solving Eqn. \eqref{equ:elliptic_relaxation} is given by performing an explicit convolution via Eqn. \eqref{equ:gaussian}, i.e., approximating $\tilde{\varphi}(x_k) \approx \int_\mathrm{\tilde{x}_\mathrm{i}} ( 0.5 \sqrt{\pi} h)^{-D} e^{-4 \left( (x_\mathrm{k} - \tilde{x}_k)/h \right)^2} \mathrm{d} \tilde{x}_i$. As shown by \cite{hojjat2014vertex}), the results obtained from implicit Sobolev smoothing and explicit kernel filtering are typically nearly identical. 

    %
    %
    %
%
    %

The paper's remainder takes explicit sources of the adjoint Eqns. \eqref{equ:adjoint_momentum_balance}-\eqref{equ:adjoint_vof_balance} as an input, e.g., $\varphi \gets \hat{c} \nabla_\mathrm{1} c$ from $r_\mathrm{1}^\mathrm{\hat{v}}$; subsequently, the output replaces their r.h.s. contribution, i.e., $r_\mathrm{1}^\mathrm{\hat{v}} \gets \tilde{\varphi}$. Vectorized expressions call the relaxation module several times, utilizing the same filter-width per direction. 

\section{Verification \& Validation}
\label{sec:validation}

This section evaluates the reliability and impact of the proposed elliptic relaxation approach. The analysis focuses on the laminar two-phase flow around a two-dimensional, submerged circular cylinder at rest, using a surface-based drag cost functional. Specifically, the functional refers to the total drag force $j^\mathrm{\Gamma} = (p \delta_\mathrm{ik} - \mu 2 S_\mathrm{ik}) n_\mathrm{k} \delta_\mathrm{i1}$ where $n_\mathrm{k}$ represents the local boundary normal, and $\delta_\mathrm{ik}$ is the Kronecker delta. A lift companion follows from $\delta_\mathrm{i1} \to \delta_\mathrm{i2}$.

\begin{figure}[!ht]
\centering
\subfigure[]{
\iftoggle{tikzExternal}{
\input{./tikz/2D_cylinder_scetch.tikz}
}{
\includegraphics{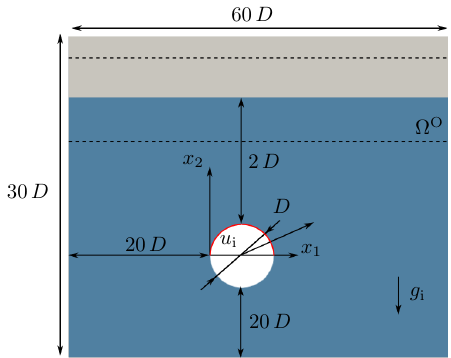}
}
}
\subfigure[]{
\includegraphics[scale=1]{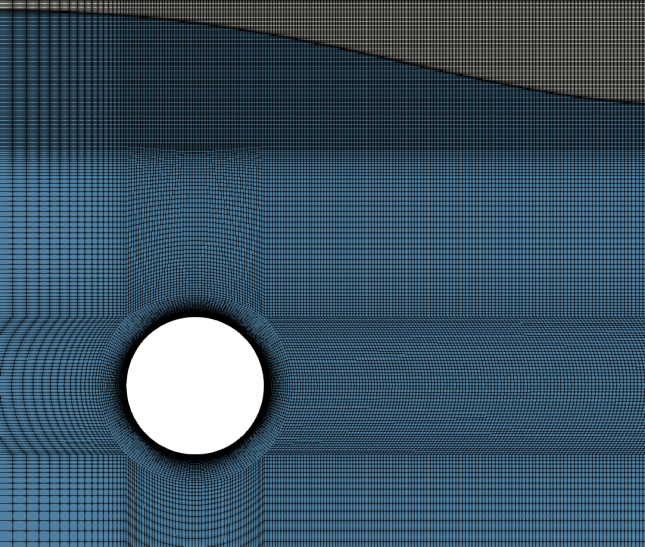}
}
\caption{Submerged cylinder case ($\mathrm{Re}_\mathrm{D} = 20$, $\mathrm{Fn}=0.75$): (a) Schematic of the initial configuration showing the controlled cylinder shape with $u_\mathrm{i}$ (red), and (b) a segment of the structured numerical grid surrounding the cylinder.}
\label{fig:cylinder_fn_075}
\end{figure}

Figure \ref{fig:cylinder_fn_075} (a) illustrates the computational setup, where the cylinder's origin is located 2.5 diameters ($D$) below the initial calm-water free surface. The two-dimensional computational domain spans a length of $60\,D$ and a height of $30\,D$, with the inlet and bottom boundaries positioned $20.5\,D$ from the cylinder's origin. A uniform horizontal bulk flow, defined as $v_\mathrm{i} = v_\mathrm{1} \delta_\mathrm{i1}$, is prescribed at the inlet for both phases, along with a calm-water concentration distribution. Slip wall conditions are applied at the top and bottom boundaries, while a hydrostatic pressure boundary condition is enforced at the outlet. To mitigate the outlet wave field and satisfy the outlet condition, the grid is stretched longitudinally in the $x_\mathrm{1}$ direction near the outlet. The study is conducted at a Reynolds number $\mathrm{Re}_\mathrm{D} = v_\mathrm{1} D / \nu_\mathrm{b} = \SI{20}{}$ and a Froude number $\mathrm{Fn} = v_\mathrm{1} / \sqrt{G \, 2 \, D} = \SI{0.75}{}$, where $G$ refers to the gravitational acceleration, $v_\mathrm{1}$ is the inflow velocity, and $\nu_\mathrm{b}$ is the kinematic viscosity of water. Based on these parameters, the expected dimensionless wavelength is $\lambda / D = 2 \pi \mathrm{Fn}^2 = 3.534$. A grid study was conducted before the paper's investigations to ensure that the objective functional is independent of spatial discretization. Fig. \ref{fig:cylinder_fn_075} (b) shows a portion of the structured numerical grid. The grid consists of approximately $\SI{215000}{}$ control volumes, with the cylinder surface discretized using 500 elements along its circumference. The non-dimensional wall-normal distance of the first grid layer is $y^+ \approx \SI{0.01}{}$. The grid is refined in the free surface region with isotropic spacing of $\Delta x_\mathrm{1} = \Delta x_\mathrm{2} \approx \lambda / 100$.

The section's focus is placed on the predictive accuracy of the elliptically relaxed adjoint-based cylinder shape sensitivities ($\mu_\varphi > 0$) compared to the results of a fully consistent adjoint study ($\mu_\varphi = 0$). The control is limited to the upper half of the cylinder (see Fig. \ref{fig:cylinder_fn_075}(a)). A direct comparison between adjoint sensitivities and Finite-Differences (FD) is omitted to save space; interested readers are referred to previous publications considering this specific two-phase cylinder flow such as \cite{kroger2018adjoint, kuhl2021adjoint, kuhl2022adjoint}. Studies therein validate the credibility of the utilized primal/dual flow solver in consistently predicting adjoint-based sensitivities for $\mu_\varphi = 0$ against FD results, which additionally were verified for linearity using three control perturbation magnitudes. 

\begin{figure}[!ht]
\centering
\iftoggle{tikzExternal}{
\input{./tikz/2D_cylinder_cinr_impression.tikz}
}{
\includegraphics{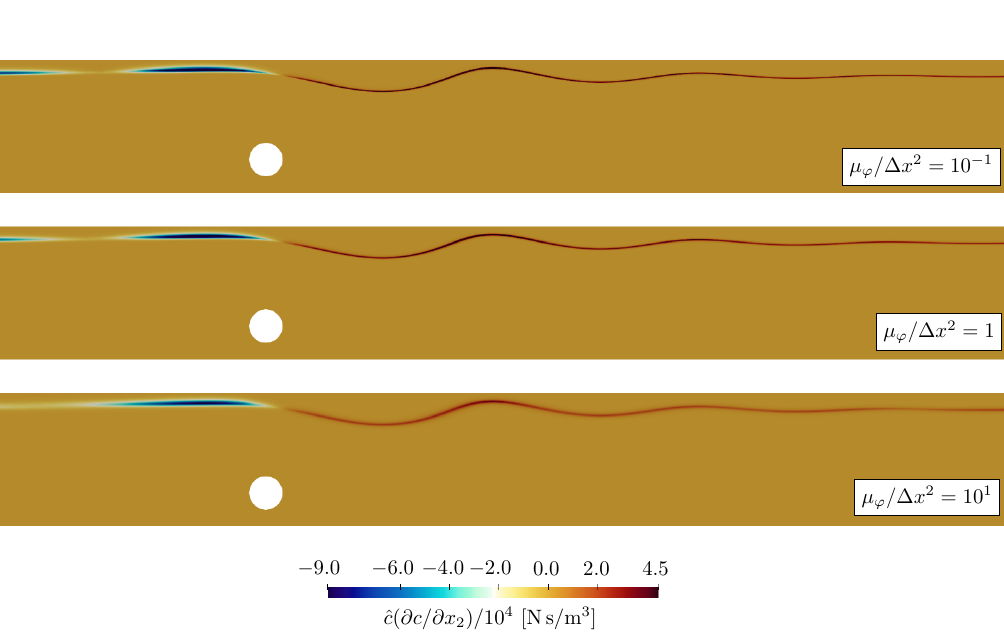}
}
\caption{Submerged cylinder case ($\mathrm{Re}_\mathrm{D} = 20$, $\mathrm{Fn}=0.75$): Converged explicit volume fraction convection contribution to the $x_2$ adjoint momentum equation for an inhomogeneous filter width of $\mu_\varphi / \Delta x^2 = [10^{-1}, 1, 10]$ from top to bottom, respectively.}
\label{fig:cylinder_fn_075_cinr}
\end{figure}

In the following, six variations of explicit source terms within the adjoint equations are elliptically relaxed: 
\begin{itemize}
   \item [(a)] The gravitational ($\rho_\Delta \hat{v}_\mathrm{i} g_i$),
   \item [(b)] the momentum convection ($\rho_\Delta \hat{v}_\mathrm{i} v_\mathrm{k} \nabla_\mathrm{k} v_i$), and 
   \item [(c)] momentum diffusion ($\mu_\Delta 2 S_\mathrm{ik} \nabla_\mathrm{k} \hat{v}_i$) contribution within the adjoint concentration Eqn. \eqref{equ:adjoint_vof_balance}, the
   \item [(d)] ATC term ($\rho \hat{v}_\mathrm{k} \nabla_\mathrm{i} v_k$) as well as the
   \item [(e)] volume fraction convection ($\hat{c} \nabla_\mathrm{i} c$) contribution to the adjoint momentum Eqns. \eqref{equ:adjoint_momentum_balance}, and
   \item [(f)] all of the former together.
\end{itemize}
Six relaxation variants are considered, which in-homogeneously relate the filter width to local grid dimensions $\mu_\varphi/\Delta x^2 = [10^{-1}, 1, 10^1]$ or homogeneously to the global cylinder diameter $\mu_\varphi/D^2 = [10^{-1}, 1, 10^1]$. The following Figs. \ref{fig:cylinder_drag_1}-\ref{fig:cylinder_drag_2} present the resulting shape sensitivities along the upper cylinder surface for relaxation combination (a)-(c) as well as (e)-(f), respectively.
\begin{figure}[!htb]
\centering
\iftoggle{tikzExternal}{
\input{./tikz/2D_cylinder_sensitivity_drag_1.tikz}}{
\includegraphics{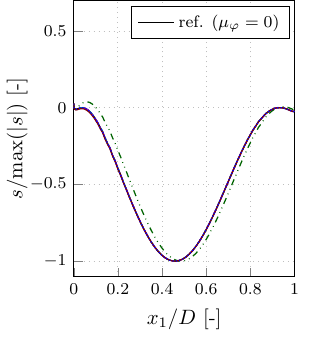}
\includegraphics{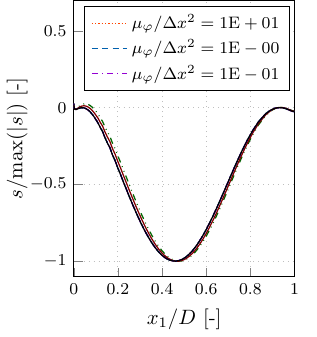}
\includegraphics{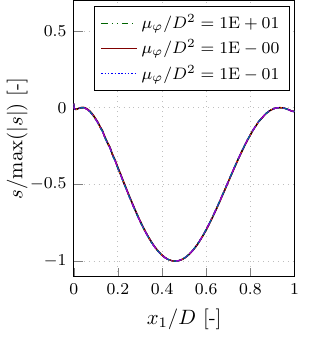}
}
\caption{Submerged cylinder case ($\mathrm{Re}_\mathrm{D} = 20$, $\mathrm{Fn}=0.75$): Normalized shape sensitivity of a drag functional along the normalized upper cylinder contour for seven different elliptic relaxation strategies with varying filter width ($\mu_\varphi/\Delta x = [0, 10^{-1}, 1, 10^1]$, $\mu_\varphi/D = [10^{-1}, 1, 10^1]$) applied to (left) the gravitational, (center) momentum convection, and (right) momentum diffusion contribution to the right-hand side of the adjoint concentration balance.}
\label{fig:cylinder_drag_1}
\end{figure}
\begin{figure}[!htb]
\centering
\iftoggle{tikzExternal}{
\input{./tikz/2D_cylinder_sensitivity_drag_2.tikz}}{
\includegraphics{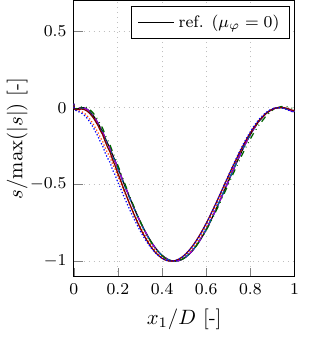}
\includegraphics{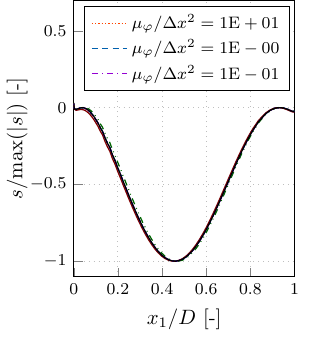}
\includegraphics{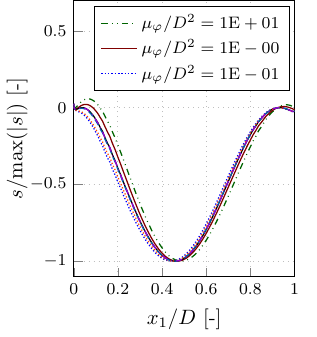}
}
\caption{Submerged cylinder case ($\mathrm{Re}_\mathrm{D} = 20$, $\mathrm{Fn}=0.75$): Normalized shape sensitivity of a drag functional along the normalized upper cylinder contour for seven different elliptic relaxation strategies with varying filter width ($\mu_\varphi/\Delta x = [0, 10^{-1}, 1, 10^1]$, $\mu_\varphi/D = [10^{-1}, 1, 10^1]$) applied to (left) the momentum and (center) volume fraction convection contribution to the right-hand side of the adjoint momentum equations as well as (right) all explicit sources to the adjoint momentum and concentration equations.}
\label{fig:cylinder_drag_2}
\end{figure}
All sensitivities are normalized based on their maximum absolute value to anticipate step size influences of a possibly subsequently utilized gradient-based optimization.
In line with previous studies on this flow (cf. \cite{kroger2018adjoint, kuhl2021continuous}), influences of adjoint concentration sources scaling with density differences are superior for the considered drag functional. Relaxing the gravity contribution provides visible deviations w.r.t. the reference data for large global filter width (cf. Fig \ref{fig:cylinder_drag_1} left) but seems to be less sensitive to all other considered relaxation strategies than the momentum convection companion (cf. Fig \ref{fig:cylinder_drag_1} center). Relaxing the momentum diffusion contribution to the adjoint concentration equation results in a minor sensitivity derivative impact (cf. Fig \ref{fig:cylinder_drag_1} right).
Influences of relaxing only adjoint momentum sources are partly higher and still do not deviate significantly from the reference results, where the ATC contribution (cf. Fig \ref{fig:cylinder_drag_2} left) is slightly more relaxation sensitive than the free surface (cf. Fig \ref{fig:cylinder_drag_2} center) contribution.
Note that the adjoint momentum source arising from the primal convective concentration transport features severe local gradients via $\hat{c} \nabla_\mathrm{i} c$ (cf. Eqn. \eqref{equ:adjoint_momentum_balance}) due to the desired as sharp as possible free surface or concentration field. Figure \ref{fig:cylinder_fn_075_cinr} provides impressions for the effect of inhomogeneously relaxing source term combination (e), i.e., the volume fraction convection contribution to the $x_2$ adjoint momentum equations.
Highest deviations are observed when relaxing all explicit sources (cf. Fig \ref{fig:cylinder_drag_2} right), especially for filter widths in the order of the cylinder diameter, whereby both local over and undershoots w.r.t. to the reference results occur. In particular, the largest considered filter width of $\mu_ \varphi/D^2 = 1$ results in shifted roots and thus sensitivities with different signs. Nevertheless, in most cases, differences in the predicted sensitivities are barely recognizable. In order to further quantify relaxation influences, relative errors of the relaxed sensitivities compared to the reference values are shown in Figs. \ref{fig:cylinder_drag_error_1}-\ref{fig:cylinder_drag_error_2}, whereby the order of relaxation combinations is consistent with Figs. \ref{fig:cylinder_drag_1}-\ref{fig:cylinder_drag_2}.
\begin{figure}[!htb]
\centering
\iftoggle{tikzExternal}{
\input{./tikz/2D_cylinder_sensitivity_drag_error_1.tikz}}{
\includegraphics{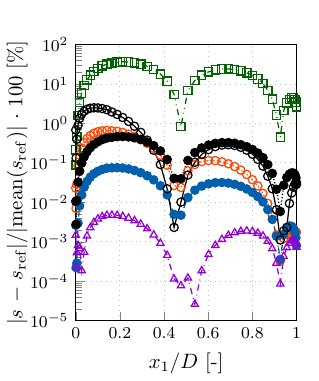}
\includegraphics{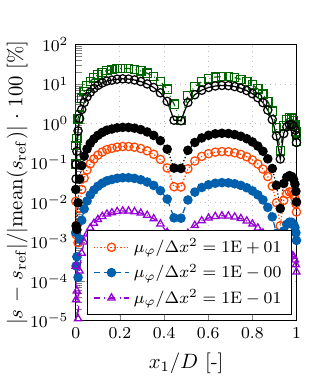}
\includegraphics{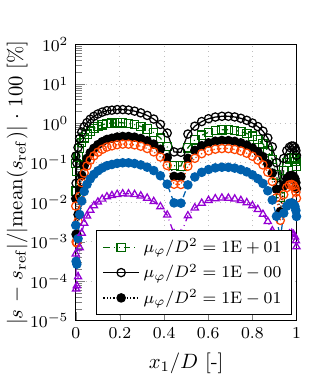}
}
\caption{Submerged cylinder case ($\mathrm{Re}_\mathrm{D} = 20$, $\mathrm{Fn}=0.75$): Relative error of the upper half cylinder's normalized drag sensitivities for the six elliptic relaxation-based results against the reference data from Figure \ref{fig:cylinder_drag_1}.}
\label{fig:cylinder_drag_error_1}
\end{figure}
\begin{figure}[!htb]
\centering
\iftoggle{tikzExternal}{
\input{./tikz/2D_cylinder_sensitivity_drag_error_2.tikz}}{
\includegraphics{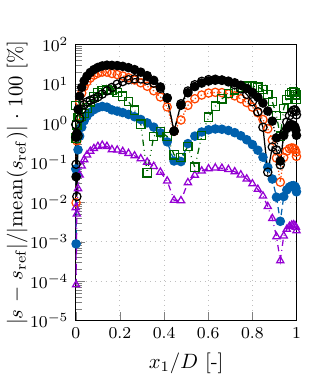}
\includegraphics{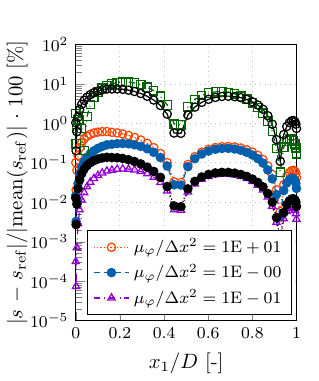}
\includegraphics{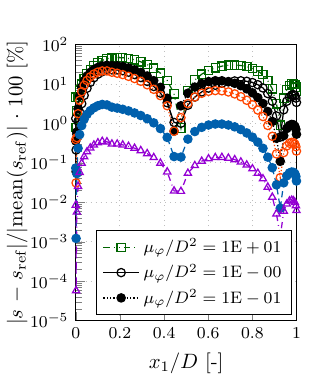}
}
\caption{Submerged cylinder case ($\mathrm{Re}_\mathrm{D} = 20$, $\mathrm{Fn}=0.75$): Relative error of the upper half cylinder's normalized drag sensitivities for the six elliptic relaxation-based results against the reference data from Figure \ref{fig:cylinder_drag_2}.}
\label{fig:cylinder_drag_error_2}
\end{figure}
In all cases, larger filter widths lead to higher deviations of 1 percent to 1 per mille for filter widths coupled to local length dimension. Error-values of the homogeneous filter widths span between 0.1 to 30 percent and feature a particular impact on contributions to the adjoint concentration equation scaling with density difference (cf. Fig \ref{fig:cylinder_drag_error_1} left). In general, due to the chosen discretization, errors of $\mu_\varphi/\Delta x^2 = 10^1$ and $\mu_\varphi/D^2 = 10^{-1}$ are approximately of similar order, and inhomogeneous local filter widths feature more minor influences compared to their homogeneous global companion.

\begin{figure}[!htb]
\centering
\iftoggle{tikzExternal}{
\input{./tikz/2D_cylinder_conservativity_drag.tikz}}{
\includegraphics{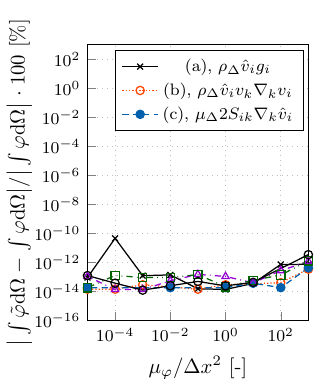}
\includegraphics{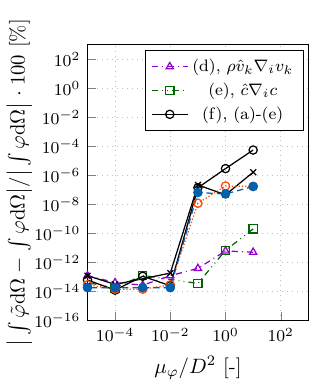}
}
\caption{Submerged cylinder case ($\mathrm{Re}_\mathrm{D} = 20$, $\mathrm{Fn}=0.75$): Conservativeness measure of all relaxation exercises over varying (left) inhomogeneous and (right) homogeneous filter widths for a drag functional.}
\label{fig:cylinder_conservativity_drag}
\end{figure}

Next, the filter width variation of all cases (a)-(f) is further escalated to smaller and larger values, i.e., $10^{-6} \leq \mu_\varphi / \Delta x^2 \leq 10^3$ and $10^{-6} \leq \mu_\varphi / D^2 \leq 10^1$. Results of the resulting averaged error in the shape derivative (see Figs. \ref{fig:cylinder_drag_1}-\ref{fig:cylinder_drag_error_2}) are shown in Fig. \ref{fig:cylinder_conservativity_drag} (left) and (right) logarithmically over the inhomogeneous and homogeneous filter width, respectively, whereby smaller [larger] filter widths lead to more minor [more significant] averaged errors. For the same filter width, errors of relaxing the ATC are around one order of magnitude larger than relaxing adjoint concentration sources, whereas relaxing the free surface contribution to the adjoint momentum equations is in between.
\begin{figure}[!htb]
\centering
\iftoggle{tikzExternal}{
\input{./tikz/2D_cylinder_error_range_drag.tikz}}{
\includegraphics{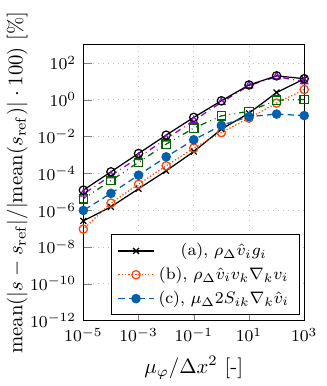}
\includegraphics{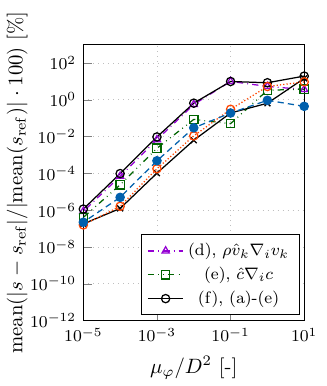}

}
\caption{Submerged cylinder case ($\mathrm{Re}_\mathrm{D} = 20$, $\mathrm{Fn}=0.75$): Mean relative error of all relaxation exercises over varying (left) inhomogeneous and (right) homogeneous filter widths for a drag functional.}
\label{fig:cylinder_error_range_drag}
\end{figure}
A relative global conservativeness measure $\big|\int \tilde{\varphi} \mathrm{d} \Omega - \int \varphi \mathrm{d} \Omega\big|/\big|\int \varphi \mathrm{d} \Omega\big| \cdot 100$ is evaluated on the relaxed expressions (a)-(f) of the converged adjoint flow fields. Results are provided in Fig. \ref{fig:cylinder_conservativity_drag} for (left) inhomogeneous and (right) homogeneous filter widths and are primarily in regions of numerical inaccuracies of the flow solver's double precision metric, underlining the global conservative character of the elliptic relaxation approach. Only global filter widths in the order of the cylinder diameter that feature fundamental kernel solutions truncated by the domain boundaries start introducing a certain degree of loss of preservation that is still in the order of $\mathcal{O}(10^{-06})$ and thus, however, still acceptable for many applications.

\begin{figure}[!htb]
\centering
\iftoggle{tikzExternal}{
\input{./tikz/2D_cylinder_sensitivity_lift_1.tikz}}{
\includegraphics{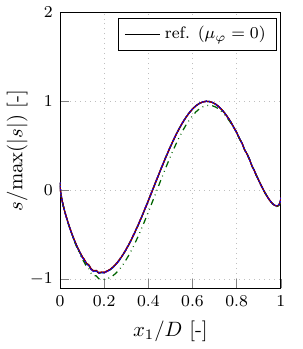}
\includegraphics{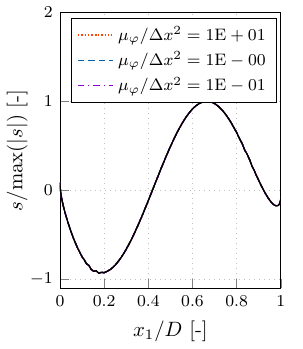}
\includegraphics{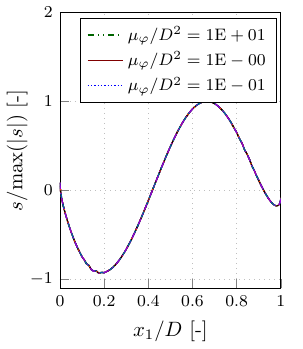}
}
\caption{Submerged cylinder case ($\mathrm{Re}_\mathrm{D} = 20$, $\mathrm{Fn}=0.75$): Normalized shape sensitivity of a lift functional along the normalized upper cylinder contour for seven different elliptic relaxation strategies with varying filter width ($\mu_\varphi/\Delta x = [0, 10^{-1}, 1, 10^1]$, $\mu_\varphi/D = [10^{-1}, 1, 10^1]$) applied to (left) the gravitational, (center) momentum convection, and (right) momentum diffusion contribution to the right-hand side of the adjoint concentration balance.}
\label{fig:cylinder_lift_1}
\end{figure}

\begin{figure}[!htb]
\centering
\iftoggle{tikzExternal}{
\input{./tikz/2D_cylinder_sensitivity_lift_2.tikz}}{
\includegraphics{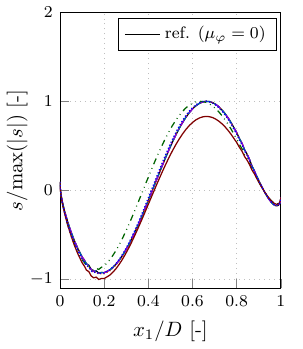}
\includegraphics{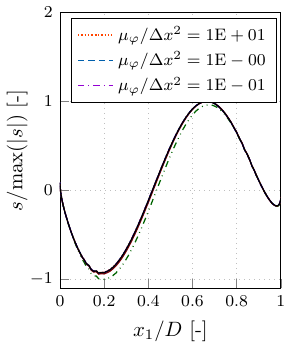}
\includegraphics{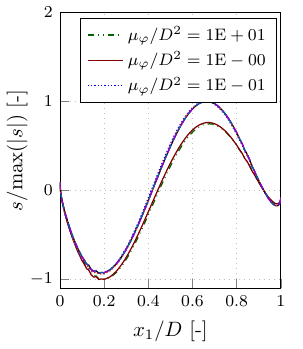}
}
\caption{Submerged cylinder case ($\mathrm{Re}_\mathrm{D} = 20$, $\mathrm{Fn}=0.75$): Normalized shape sensitivity of a lift functional along the normalized upper cylinder contour for seven different elliptic relaxation strategies with varying filter width ($\mu_\varphi/\Delta x = [0, 10^{-1}, 1, 10^1]$, $\mu_\varphi/D = [10^{-1}, 1, 10^1]$) applied to (left) the momentum and (center) volume fraction convection contribution to the right-hand side of the adjoint momentum equations as well as (right) all explicit sources to the adjoint momentum and concentration equations.}
\label{fig:cylinder_lift_2}
\end{figure}

\begin{figure}[!htb]
\centering
\iftoggle{tikzExternal}{
\input{./tikz/2D_cylinder_sensitivity_lift_error_1.tikz}}{
\includegraphics{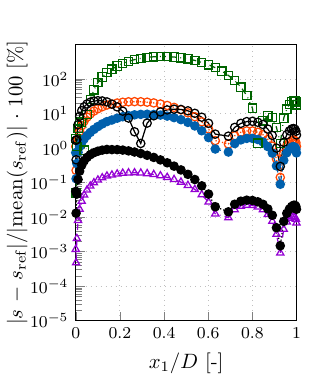}
\includegraphics{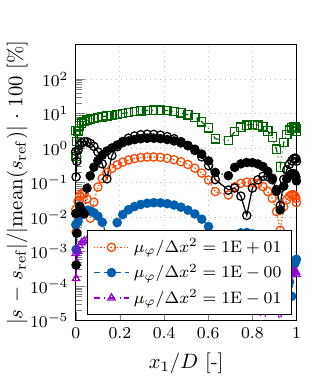}
\includegraphics{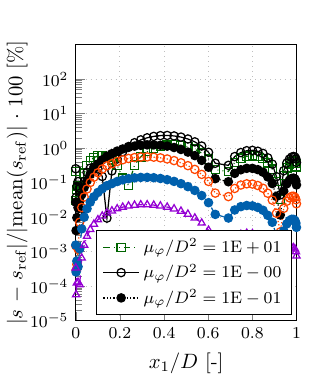}
}
\caption{Submerged cylinder case ($\mathrm{Re}_\mathrm{D} = 20$, $\mathrm{Fn}=0.75$): Relative error of the upper half cylinder's normalized lift sensitivities for the six elliptic relaxation-based results against the reference data from Figure \ref{fig:cylinder_lift_1}.}
\label{fig:cylinder_lift_error_1}
\end{figure}

\begin{figure}[!htb]
\centering
\iftoggle{tikzExternal}{
\input{./tikz/2D_cylinder_sensitivity_lift_error_2.tikz}}{
\includegraphics{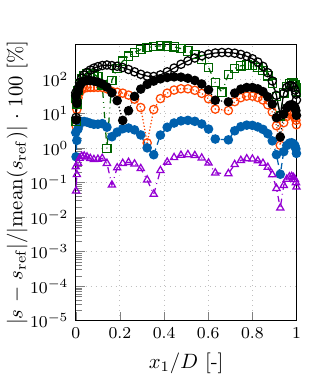}
\includegraphics{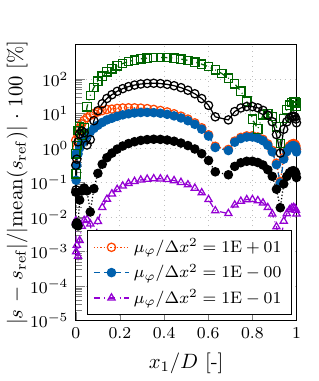}
\includegraphics{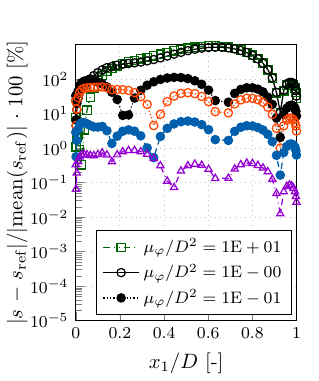}
}
\caption{Submerged cylinder case ($\mathrm{Re}_\mathrm{D} = 20$, $\mathrm{Fn}=0.75$): Relative error of the upper half cylinder's normalized lift sensitivities for the six elliptic relaxation-based results against the reference data from Figure \ref{fig:cylinder_lift_2}.}
\label{fig:cylinder_lift_error_2}
\end{figure}

\begin{figure}[!htb]
\centering
\iftoggle{tikzExternal}{
\input{./tikz/2D_cylinder_error_range_lift.tikz}}{
\includegraphics{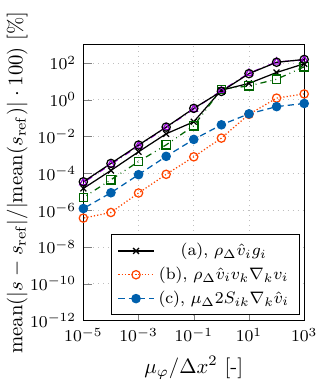}
\includegraphics{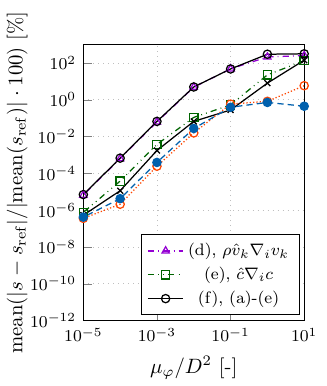}
}
\caption{Submerged cylinder case ($\mathrm{Re}_\mathrm{D} = 20$, $\mathrm{Fn}=0.75$): Mean relative error of all relaxation exercises over varying (left) inhomogeneous and (right) homogeneous filter widths for a lift functional.}
\label{fig:cylinder_error_range_lift}
\end{figure}

\begin{figure}[!htb]
\centering
\iftoggle{tikzExternal}{
\input{./tikz/2D_cylinder_conservativity_lift.tikz}}{
\includegraphics{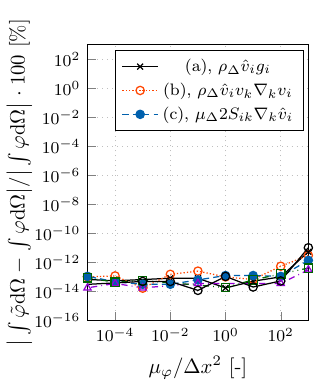}
\includegraphics{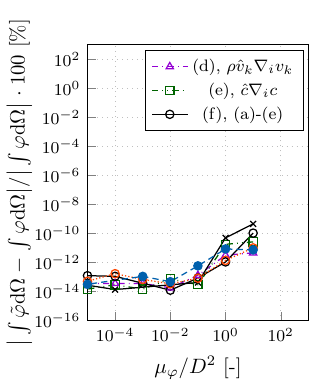}
}
\caption{Submerged cylinder case ($\mathrm{Re}_\mathrm{D} = 20$, $\mathrm{Fn}=0.75$): Conservativeness measure of all relaxation exercises over varying (left) inhomogeneous and (right) homogeneous filter widths for a lift functional.}
\label{fig:cylinder_conservativity_lift}
\end{figure}

The exercise is performed analogously for a lift cost functional. Normalized sensitivity results and relative error curves are shown in Figs. \ref{fig:cylinder_lift_1}-\ref{fig:cylinder_conservativity_lift}, whereby the figure arrangement is consistent with the previous investigation of the resistance functional. The influence of relaxing the convective momentum source for the adjoint concentration equation is reduced, and a relaxation of the ATC term now exhibits more significant deviations and errors than considering the gravity contribution. However, results are mostly congruent with those of the drag function; in particular, the relative errors for local filter widths $\mu_\varphi/\Delta x^2 = 1$ are again in single-digit percentage order of magnitude.

It should be emphasized that for the considered objective functions, an inhomogeneous filter width in the order of the local grid spacing (i.e. $\mu_\varphi / \Delta x = \mathcal{O}(10)$) leads to single-digit percentage mean errors (i.e. $\mathrm{mean}(|s - s_\mathrm{ref}|/|\mathrm{mean}(s_\mathrm{ref})|) = \mathcal{O}(1\%)$), that motivates applying the proposed elliptical relaxation strategies during industrial adjoint-based applications.

\section{Application}
\label{sec:application}

In the following, the presented elliptic relaxation strategy is applied for the numerical stabilization of the adjoint simulation during the hull optimizations of a bulk carrier and a harbor ferry. The optimizations address the reduction of the total resistance, i.e., the cost functional corresponds to the drag functional from Eqn. \eqref{equ:general_objective}, cf. beginning of Sec. \ref{sec:validation}. While the bulk carrier is investigated at a comparatively small Froude number of Fn=0.142 at model scale, the optimization of the harbor ferry is carried out at a significantly higher Froude number of Fn=0.4 in full-scale conditions. The second test case is particularly challenging due to its unfavorable (hump-near) Froude number, the breaking bow wave, the substantial change in the floating position, and the full-scale Reynolds number.

The shape optimizations utilize a parameter-free Steklov-Poincare shape metric, which uses the sensitivity from Eqn. \eqref{equ:sensitivity_derivative} as a Neumann boundary condition and calculates a deformation field in the discretized volume. This deformation field is multiplied by a small step-size value in line with a gradient-descent method and then applied to the present numerical grid. Subsequently, the cost function is analyzed again, i.e., the primal resistance simulation is repeated, whereby restart effects permit shorter simulation times. The optimization process is terminated if the change in the objective function is negligible.

In order to assess the quality of the approximated shape derivatives and, in particular, the influence of the elliptical relaxation method, the number of technical constraints is kept as low as possible by keeping only (a) the volume of water displaced by the hull constant and (b) the adherence to a plane transom that allows only for tangential deformation. The step size of the gradient method is selected so that the maximum deformation of the initial shape update refers to a fraction of the hull length, e.g., one per mil of the length between the perpendiculars ($L_\mathrm{pp}$).

To fundamentally increase the stability of the adjoint system, the adjoint Eqns. \eqref{equ:adjoint_mass_balance}-\eqref{equ:adjoint_vof_balance} are integrated within (a) a pseudo-time (\cite{kuhl2021continuous}) and (b) approximated with solid under-relaxation parameters that refer to less than half of those employed for the primal solution process. Further stability-enhancing measures are refer to (c) an adjoint pressure-velocity coupling procedure that is as discretely consistent as possible (\cite{kuhl2022discrete} and (d) the use of artificial viscosities (\cite{kuhl2021continuous}) within the adjoint concentration equation. As the results of this chapter will show, these stabilizing measures are generally insufficient to solve consistent two-phase adjoint systems robustly on industrial grids.

\subsection{Bulk Carrier -- Low Froude-Number Case}
\label{subsec:jbc}
\begin{figure}[!h]
\centering
\subfigure[]{
\iftoggle{tikzExternal}{
\input{./tikz/3D_jbc_scetch.tikz}}{
\includegraphics{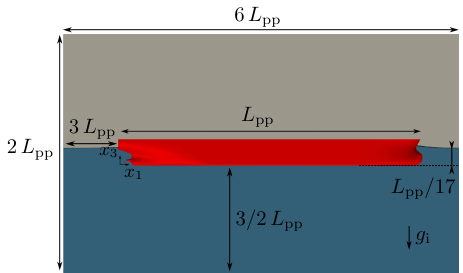}}
}
\subfigure[]{
\includegraphics[width=0.4\textwidth]{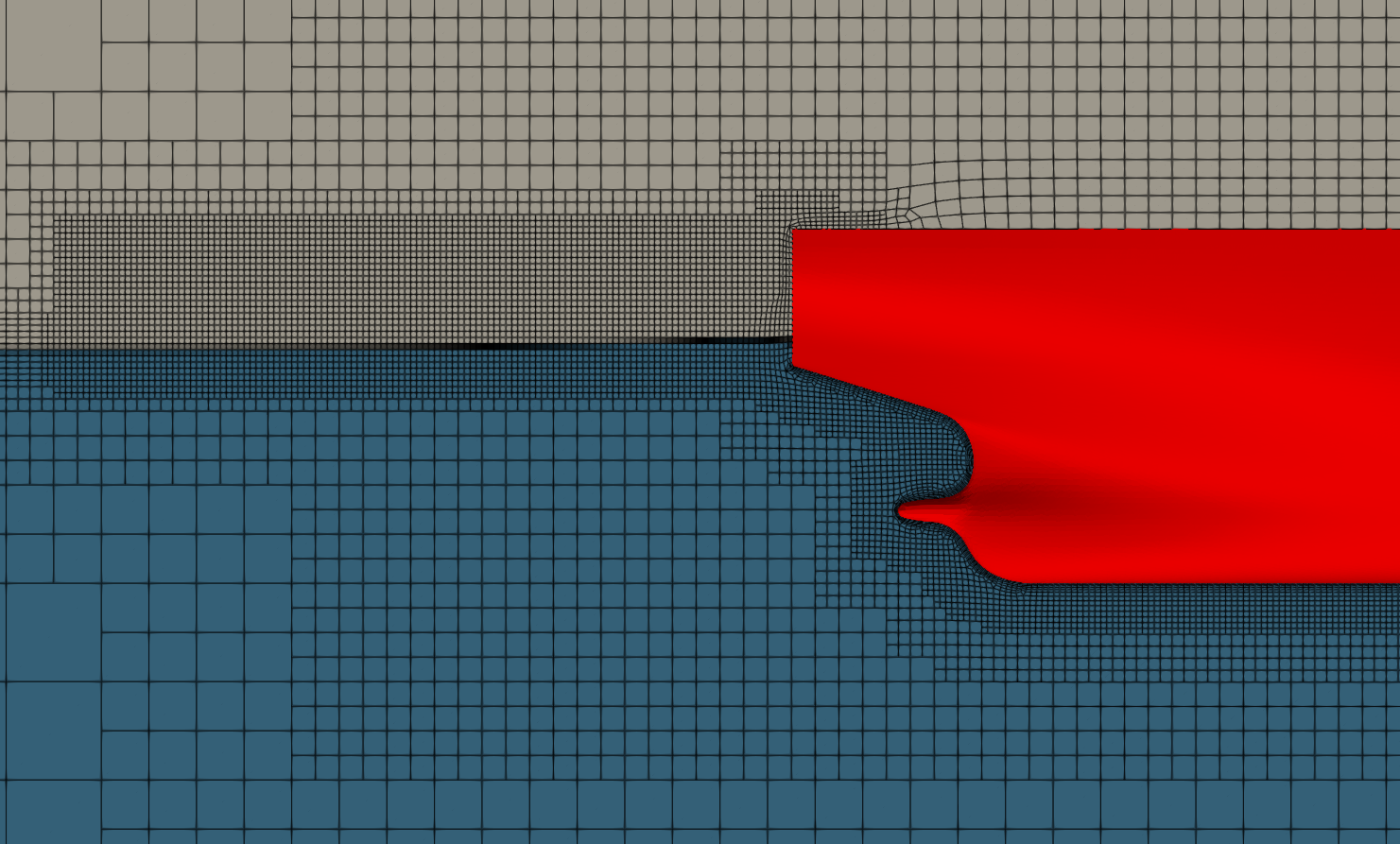}
}
\caption{Japan Bulk Carrier case ($\mathrm{Re}_\mathrm{L} = \SI{7.246}{} \cdot 10^6$, $\mathrm{Fn}=0.142$): (a) Schematic drawing of the initial configuration and (b) unstructured numerical grid around the 
stern region.}
\label{fig:jbc_scetch}
\end{figure}

The Japan Bulk Carrier (JBC) of the 2015 Tokyo CFD workshop (\cite{hino2020numerical}) is analyzed using a model of scale 40. The Reynolds and Froude numbers are $\mathrm{Re} = V L_\mathrm{pp} / \nu = \SI{7.246E+06}{}$ and $\mathrm{Fn} = V / \sqrt{L_\mathrm{pp} \, G} = 0.142$, based on the towing speed $V = \SI{1.179}{m/s}$, the length between the perpendiculars $L_\mathrm{pp} = \SI{7}{m}$, the dynamic viscosity of water $\nu = \SI{1.139E-06}{m^2/s}$, and the gravitational constant of $G = \SI{9.81}{m/s^2}$. The calm water resistance studies feature no hull attachments, particularly no energy-saving devices. The JBC consists of a hull, transom, and deck as conceptually sketched in Fig. \ref{fig:jbc_scetch}. The analysis focuses on the steady-state conditions in calm water, with flotation effects being suppressed. As shown in Fig. \ref{fig:jbc_scetch}, the origin of the Cartesian coordinate system is positioned beneath the transom stern of the initial configuration. The free surface is initialized in the $x_\mathrm{1}-x_\mathrm{2}$ plane at a height of $x_\mathrm{3} / L_\mathrm{pp} = 1 / 17$.
The simulation domain has dimensions of $6 \, L_\mathrm{pp}$ in length, $2 \, L_\mathrm{pp}$ height, and $2 \, L_\mathrm{pp}$ in width. The outlet and bottom boundaries are positioned at distances of three and three-and-a-half hull lengths, respectively, from the origin. The expected dimensionless wavelength is $\lambda / L_\mathrm{pp} = 2 \, \pi \, \mathrm{Fn}^2 = 0.852$.

\begin{figure}[!h]
\centering
\includegraphics[width=0.9\textwidth]{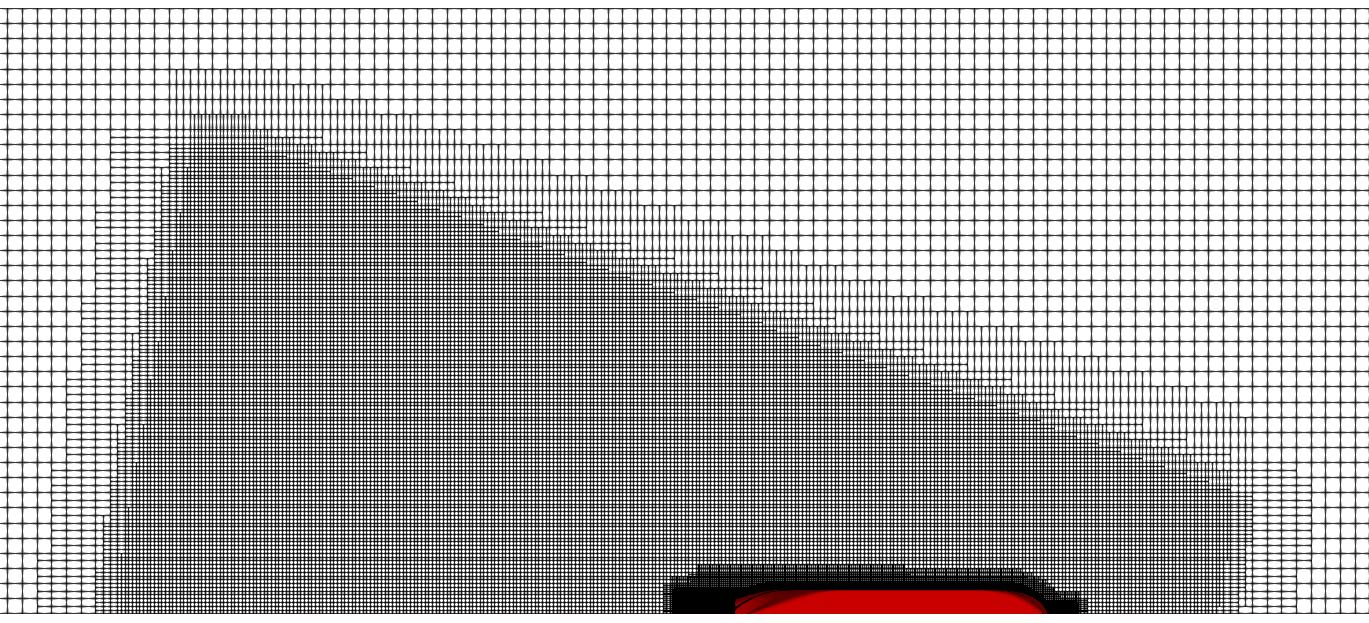}
\caption{Japan Bulk Carrier case ($\mathrm{Re}_\mathrm{L} = \SI{7.246}{} \cdot 10^6$, $\mathrm{Fn}=0.142$): Numerical grid in the still water plane.}
\label{fig:jbc_free_surface}
\end{figure}

The unstructured numerical grid comprises approximately $\SI{2.5}{} \cdot 10^6$ control volumes, and an impression of the grid around the transom is provided in Fig. \ref{fig:jbc_scetch} (b). Due to symmetry, only half of the geometry is modeled in the lateral ($x_\mathrm{2}$) direction. The simulations are fully turbulent and utilize the wall-function-based $k-\omega$ SST model from \cite{menter1994two}, with a non-dimensional wall-normal distance of $y^+ \approx \SI{30}{}$ for the first grid layer next to the hull. The horizontal resolution of the free surface region is refined within a Kelvin wedge to capture the wave field generated by the vessel accurately (see Fig. \ref{fig:jbc_free_surface}). The free surface resolution corresponds to approximately $\Delta x_\mathrm{1} / \lambda = \Delta x_\mathrm{2} / \lambda = 1/50$ cells in the horizontal directions and $\Delta x_\mathrm{3} / \lambda = 1/500$ cells in the vertical direction. Convective primal and adjoint fluxes are approximated using the QUICK and QDICK schemes, respectively, as described in \cite{stuck2013adjoint}. Primal simulations are advanced to a steady-state solution in pseudo-time with Courant numbers $\mathrm{Co} \leq 0.9$, employing an Euler implicit method. A uniform horizontal bulk flow is applied along the outer boundaries for both phases, accompanied by a calm water concentration distribution. At the upper boundary, a hydrostatic pressure condition is imposed. Consistent with the validation study, the grid is stretched near the outlet to dampen the wave field and satisfy the outlet condition. Additionally, a symmetry condition is enforced along the midship plane.

\begin{figure}[!h]
\centering
\iftoggle{tikzExternal}{
\input{./tikz/3D_jbc_cinr_sens_diff.tikz}}{
\includegraphics{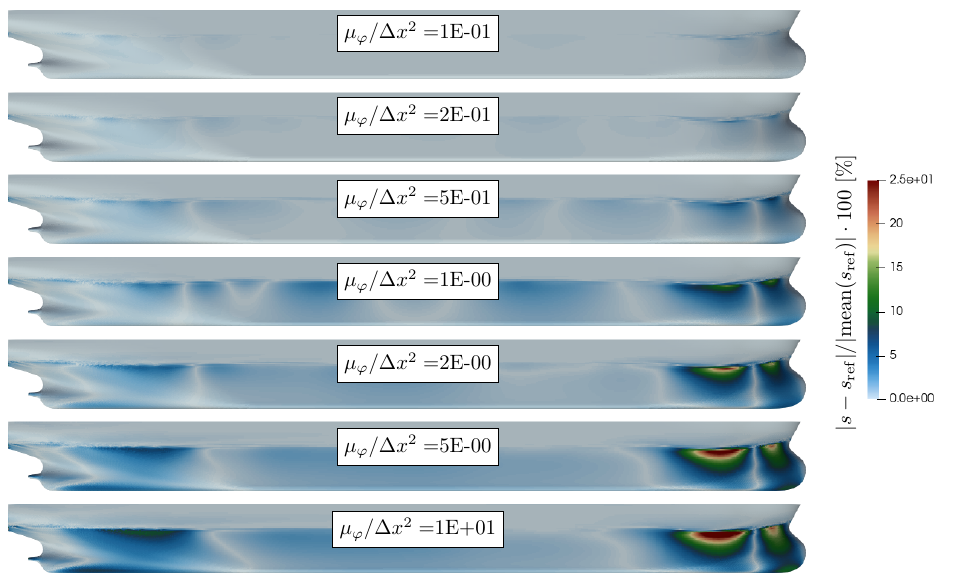}
}
\caption{Japan Bulk Carrier case ($\mathrm{Re}_\mathrm{L} = \SI{7.246}{} \cdot 10^6$, $\mathrm{Fn}=0.142$): Relative local differences in hull resistance sensitivity for differently relaxed adjoint two-phase systems against a reference sensitivity without elliptic relaxation. All sensitivities follow experiment E2 and are computed without the Adjoint Transpose Convection contribution to the adjoint momentum equations.}
\label{fig:jbc_cinr_sens_diff}
\end{figure}

\begin{figure}[!h]
\centering
\iftoggle{tikzExternal}{
\input{./tikz/3D_jbc_rinr_sens_diff.tikz}}{
\includegraphics{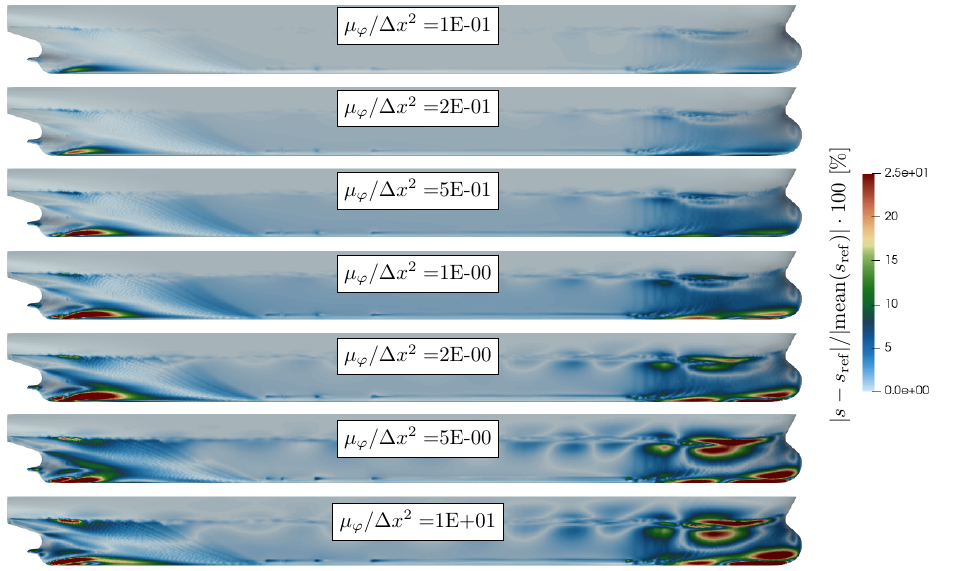}
}
\caption{Japan Bulk Carrier case ($\mathrm{Re}_\mathrm{L} = \SI{7.246}{} \cdot 10^6$, $\mathrm{Fn}=0.142$): Relative local differences in hull resistance sensitivity for differently relaxed adjoint two-phase systems against a reference sensitivity without elliptic relaxation. All sensitivities follow experiment E3 and thus a frozen free surface approach, i.e., they are computed without the free-surface contribution to the adjoint momentum equations.}
\label{fig:jbc_rinr_sens_diff}
\end{figure}

\begin{figure}[!h]
\centering
\iftoggle{tikzExternal}{
\input{./tikz/3D_jbc_consistent_sens_diff.tikz}}{
\includegraphics{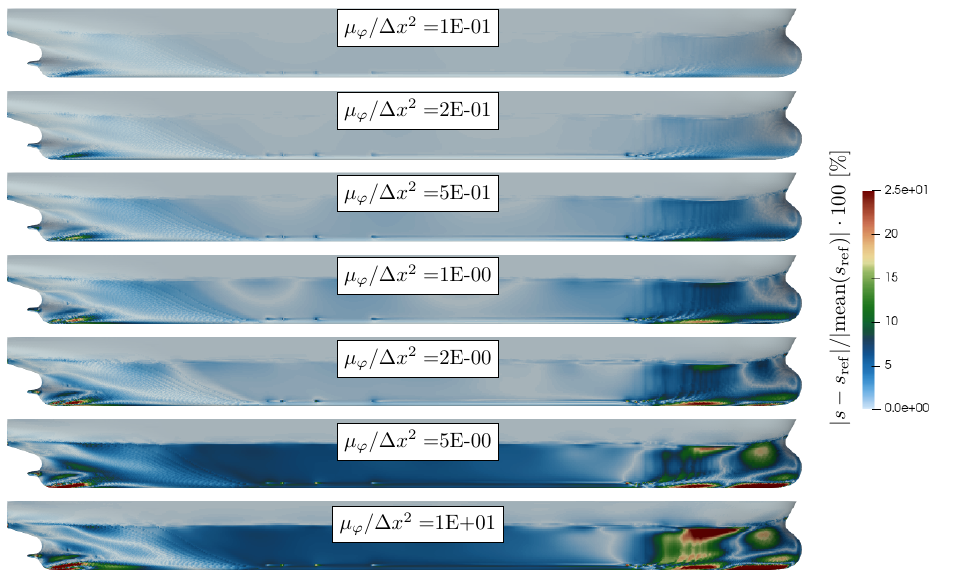}
}
\caption{Japan Bulk Carrier case ($\mathrm{Re}_\mathrm{L} = \SI{7.246}{} \cdot 10^6$, $\mathrm{Fn}=0.142$): Relative local differences in hull resistance sensitivity for differently relaxed adjoint two-phase systems against a reference sensitivity without elliptic relaxation. All sensitivities follow experiment E3 and are consistent with Eqns. \eqref{equ:adjoint_momentum_balance}-\eqref{equ:adjoint_vof_balance}.}
\label{fig:jbc_consistent_sens_diff}
\end{figure}

In the following, different numerical experiments are conducted based on four different adjoint strategies, which differ in the strength of the equation coupling and elliptic relaxation efforts:
\begin{enumerate}
    \item[E1:] All explicit cross-coupling terms on the right-hand side of the adjoint momentum equations are neglected. These are the ATC ($\rho \hat{v}_\mathrm{k} \nabla_\mathrm{i} v_k$) and the free surface ($\hat{c} \nabla_\mathrm{i} c$) contributions. Hence, this approach resembles a frozen-free-surface approach.
    \item[E2:] ATC contributions are neglected, but the free surface contributions are evaluated.
    \item[E3:] The opposite of E2, i.e., only the ATC term is considered, and the free surface contribution is neglected. Again, this refers to a frozen-free-surface approach.
    \item[E4:] All contributions to the adjoint momentum equations are considered in line with a maximally consistent adjoint system.
\end{enumerate}  
Adjoint r.h.s. contributions of experiment E2-E4 are systematically relaxed based on inhomogeneous and homogeneous filter widths scaling with the local grid spacing and the vessel's length, respectively, i.e., $\mu_\varphi / \Delta x^2 = [0, 0.1, 0.2, 0.5, 1, 2, 5, 10]$ and $\mu_\varphi / L_\mathrm{pp}^2 = [0.1, 0.2, 0.5, 1, 2, 5, 10] \cdot 10^{-3}$, resulting in $3 \times (8+7) = 45$ adjoint simulations for E2-E4. No elliptic relaxation is applied for experiment E1, as no explicit adjoint momentum sources are considered. Local relative errors in the resulting hull resistance sensitivities based on inhomogeneous filter widths against reference simulations with vanishing filter width $|s-s_\mathrm{ref}|/|\mathrm{mean}(s_\mathrm{ref})| \cdot 100$ [\%] are depicted in Figs. \ref{fig:jbc_cinr_sens_diff}-\ref{fig:jbc_consistent_sens_diff} for E2-E4, respectively. Whereas filter width in the order of $\mu_\varphi / \Delta x^2 =\mathcal{O}(0.1)$ results in local errors of $|s-s_\mathrm{ref}|/|\mathrm{mean}(s_\mathrm{ref})| \cdot 100 = \mathcal{O}(1 \%)$, an increased filter width of $\mu_\varphi / \Delta x^2 = \mathcal{O}(10)$ results in increased relative errors of $|s-s_\mathrm{ref}|/|\mathrm{mean}(s_\mathrm{ref})| \cdot 100 = \mathcal{O}(10 \%)$. In all cases, errors are minor along the mid-ship region but pronounced in the for- and aft-ship regions, where errors of experiment E2 seem smoother than errors in E3-E4. In addition, the defects in the shape derivative of E2 are only pronounced near the shoulder wave at the for-ship. In contrast, a relaxation of the ATC term in E3 also appears to cause significant changes in the fore and aft keel.

\begin{figure}[!htb]
\centering
\iftoggle{tikzExternal}{
\input{./tikz/3D_jbc_mean_error.tikz}}{
\includegraphics{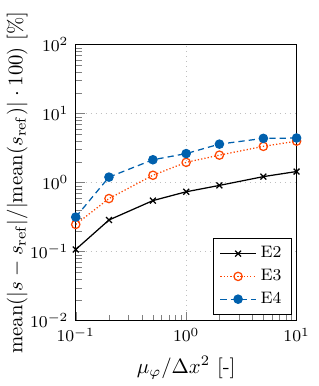}
\includegraphics{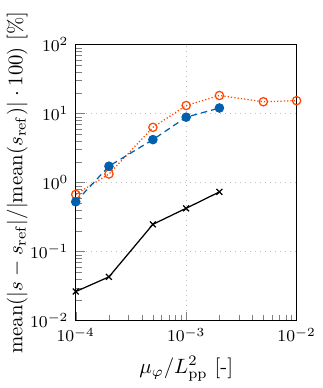}
}
\caption{Japan Bulk Carrier case ($\mathrm{Re}_\mathrm{L} = \SI{7.246}{} \cdot 10^6$, $\mathrm{Fn}=0.142$): Mean relative error of all relaxation experiments over varying (left) inhomogeneous and (right) homogeneous filter widths. Data for E2 and E4 for $\mu_\varphi / L_\mathrm{pp}^2 = [0.5,1] \cdot 10^{-2}$ is missing due to convergence issues.}
\label{fig:jbc_mean_error}
\end{figure}

Following the studies from Sec. \ref{sec:validation}, local errors in elliptically relaxed shape derivatives are averaged via $\mathrm{mean}(|s-s_\mathrm{ref}|/|\mathrm{mean}(s_\mathrm{ref})| \cdot 100)$ [\%] over the vessel's entire hull. Results of experiments E2-E4 are provided in Fig. \ref{fig:jbc_mean_error} for inhomogeneous (left) as well as homogeneous (right) filter widths. As expected from Figs. \ref{fig:jbc_cinr_sens_diff}-\ref{fig:jbc_consistent_sens_diff},  mean errors for inhomogeneous filter widths are most minor for E2 and largest for E4. Results of E3 are in between but are closer to those of E4 for inhomogeneous filter width and almost the same for homogeneous approaches. Data of the latter for $\mu_\varphi / L_\mathrm{pp}^2 = [0.5,1] \cdot 10^{-2}$ of experiment E2 and E4 is missing due to adjoint solver convergence issues. Relaxation of the free water surface contribution to the adjoint momentum equation no longer seems advantageous above a certain homogeneous filter width, as the first adjoint simulation already diverges.
As expected for such a low Froude number, the minor influence on relaxing the free-surface contribution to the adjoint momentum equations indicates a reduced contribution of the free water surface and, thus, the total resistance's wave fraction. Overall, the error range aligns with previous validation studies from Sec. \ref{sec:validation}, cf. Figs. \ref{fig:cylinder_conservativity_drag} and \ref{fig:cylinder_conservativity_lift} left.

\begin{figure}[!htb]
\centering
\iftoggle{tikzExternal}{
\input{./tikz/3D_jbc_obj_type_1.tikz}}{
\includegraphics{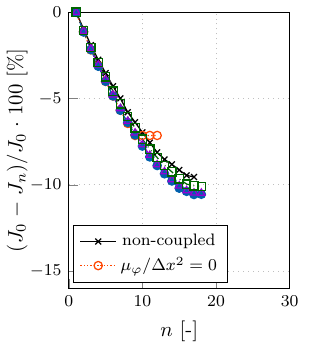}
\includegraphics{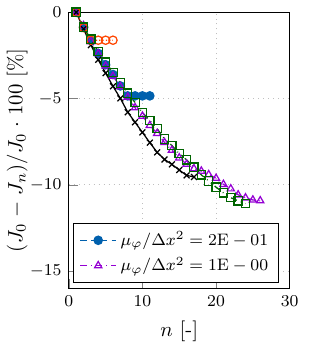}
\includegraphics{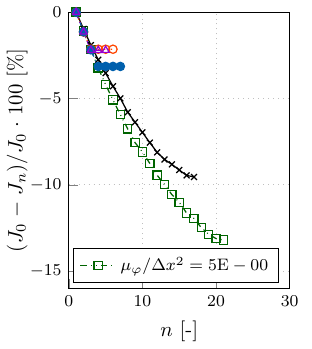}
}
\caption{Japan Bulk Carrier case ($\mathrm{Re}_\mathrm{L} = \SI{7.246}{} \cdot 10^6$, $\mathrm{Fn}=0.142$): Decrease in the objective functional over the number of optimization gradient steps for different elliptic relaxation strategies with varying inhomogeneous filter width ($\mu_\varphi/\Delta x^2 = [0, 0.2, 1, 5]$, applied in line with experiment E2 (left), E3 (center), as well as E4 (right). Convergence of optimization E1 that neglects all explicit adjoint sources is added to all graphs and serves as a benchmark.}
\label{fig:jbc_obj_type_1}
\end{figure}

\begin{figure}[!htb]
\centering
\iftoggle{tikzExternal}{
\input{./tikz/3D_jbc_obj_type_2.tikz}}{
\includegraphics{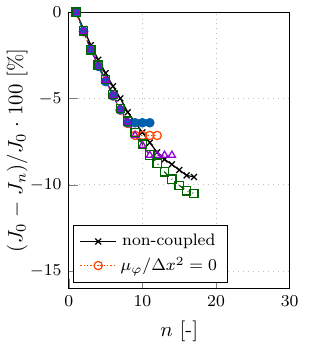}
\includegraphics{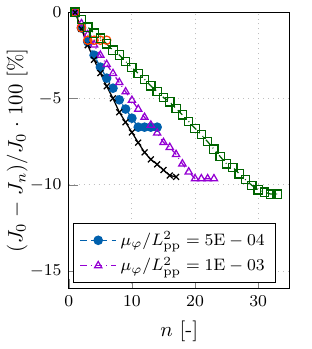}
\includegraphics{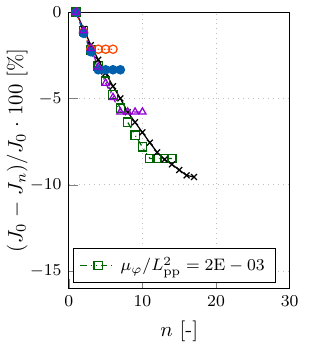}
}
\caption{Japan Bulk Carrier case ($\mathrm{Re}_\mathrm{L} = \SI{7.246}{} \cdot 10^6$, $\mathrm{Fn}=0.142$): Decrease in the objective functional over the number of optimization gradient steps for different elliptic relaxation strategies with varying filter homogeneous width ($\mu_\varphi/L_\mathrm{pp}^2 = [0, 0.5, 1, 2] \cdot 10^{-3}$, applied in line with experiment E2 (left), E3 (center), as well as E4 (right). Convergence of optimization E1 that neglects all explicit adjoint sources is added to all graphs and serves as benchmark.}
\label{fig:jbc_obj_type_2}
\end{figure}

Next, all experiments E2-E4 run through a complete shape optimization for selected filter widths $\mu_\varphi / \Delta x^2 = [0, 0.2, 1, 5]$ and $\mu_\varphi / L_\mathrm{pp}^2 = [0.5, 1. 2] \dot 10^{-3}$. Together with the single optimization of E1, a total of 22 optimization are carried out. The behavior of the relative cost functional $(J_0 - J_n)/J_0 \cdot 100$ [\%] is shown in Figs. \ref{fig:jbc_obj_type_1}-\ref{fig:jbc_obj_type_2} for inhomogeneous and homogeneous filter widths, respectively. In both figures, experiment E1 (E2) [E3] is shown on the left (center) [right]. The convergence of the optimization E1 is added to all three graphs as a reference. When an optimization terminates due to an adjoint solver divergence, the cost function is plotted with the constant last objective decrease for three more optimization steps.

First, the optimizations with locally variable filter widths of Fig. \ref{fig:jbc_obj_type_1} are discussed. In line with the previous error results from Figs. \ref{fig:jbc_cinr_sens_diff}-\ref{fig:jbc_mean_error}, the frozen ATC optimizations from experiment E2 depicted on the left are less sensitive to different filter widths, and all optimizations except one converge to similar cost functional values with a decrease of about 10\%. Although even experiment E1 ends up in this order of magnitude, the more consistent treatment leads, as expected, to increased cost functional reductions, even in the initial stage of optimization --remember that initial maximum displacement value coincides in all cases. A slight under-performance is observed for $\mu_\varphi / \Delta x^2 = 5$ compared to $\mu_\varphi / \Delta x^2 = [1, 2]$, i.e., more substantial relaxation. In the case of a vanishing relaxation $\mu_\varphi / \Delta x^2 = 0$, the optimization terminates after nine optimization steps due to divergence of the adjoint solver, which underlines the motivation of the underlying paper. Interestingly, even minimal relaxation efforts well below the grid resolution positively influence stability.
The hull's resistance sensitivity appears to be significantly more sensitive w.r.t. the ATC term, as can already be anticipated from Figs. \ref{fig:jbc_rinr_sens_diff}-\ref{fig:jbc_mean_error}. This is underlined by Fig. \ref{fig:jbc_obj_type_1} (center), in which all optimizations with a filter width below the grid resolution are unsuccessful and, therefore, finally cause a lower reduction of the cost function than the inconsistent but more robust adjoint system of experiment E1. The optimization without relaxation terminates after three gradient steps with a resistance reduction of approx. 2\%. The smallest filter width of one-fifth of the local grid spacing progresses five steps further and ends at the eighth optimization step, with a reduction of approx. 5\%. Both optimizations with filter widths equal or above the local grid resolution run through successfully without adjoint divergence issues and converge after 26 and 24 optimization steps for $\mu_\varphi / \Delta x = 1$ and $\mu_\varphi / \Delta x = 5$, respectively, to cost function reductions of approx. 11\% and thus below the frozen ATC \& free-surface approach from experiment E1. Interestingly, both successful methods converge more slowly from the beginning of the optimization, and the higher filter width --and, therefore, stronger manipulation-- leads to a slightly further drag reduction.
Results of optimizations based on E4 are shown in Fig. \ref{fig:jbc_obj_type_1} on the right and are in line with the two previous discussions, i.e., larger filter widths positively influence the numerical stability of the optimization process and thus lead to an improved resistance reduction than experiment E1. While all other consistent optimizations terminate after 3-4 gradient steps, only case $\mu_\varphi / \Delta x^2 = 5$ does not experience any adjoint solver stability issues and converges to approx. 4\% lower relative resistance values compared to E1. The resistance reduction has already increased in the initial optimization period, and convergence of the optimization method is reached after 21 gradient steps, that is, between the number of optimization steps from E2 and E3, whereby a clear out-performance of both is achieved.

Results of the optimizations with homogeneous filter widths from Fig. \ref{fig:jbc_obj_type_2} are mainly similar to those of the inhomogeneous filter widths from the previous discussion. Small filter widths tend to terminate the optimization loop prematurely; large filter widths yield less steep initial descent directions, and more consistent adjoint systems generally achieve more considerable drag reductions. In the case of both half-consistent experiments, E2 and E3, only the largest filter width can push the optimization into convergence. The latter is no longer achieved at all in the case of the entirely consistent system E4, so an optimization based on the decoupled adjoint system E1 achieves overall higher resistance reductions.

\begin{figure}[!h]
\centering
\iftoggle{tikzExternal}{
\input{./tikz/3D_jbc_wave_fields.tikz}}{
\includegraphics{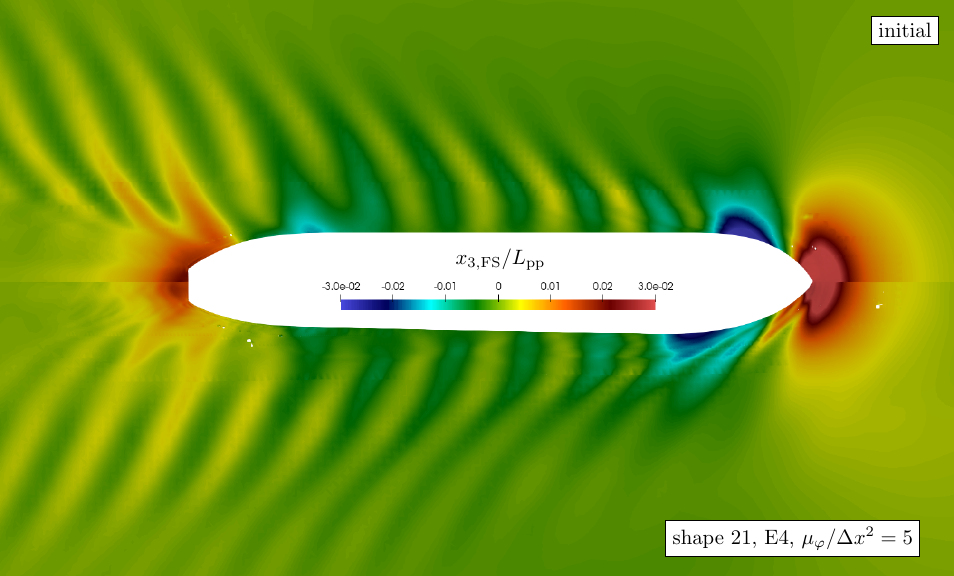}
}
\caption{Japan Bulk Carrier case ($\mathrm{Re}_\mathrm{L} = \SI{7.246}{} \cdot 10^6$, $\mathrm{Fn}=0.142$): Wave elevation of the initial (top) and the optimized (bottom) hull with the highest resistance reduction.}
\label{fig:jbc_wave_fields}
\end{figure}

The two-digit relative reduction in total resistance by approximately 14\% of the consistent optimization with an inhomogeneous filter width of $\mu_\varphi/\Delta x^2 = 5$ can be seen in the vessel's wave field. Figure \ref{fig:jbc_wave_fields} compares those of the initial (top) and optimized (bottom, shape 21 of E4) hull shape. An apparent reduction in the bow wave, the front shoulder wave, and an overall solid reduction in the stern wave system can be seen.

The calculation time of the elliptical relaxation equation is generally acceptable and only increases significantly for large globally homogeneous filter widths that increase the system's matrix bandwidth. The increase is about 10-50\% of the non-relaxed adjoint solver's simulation time and, thus, generally tolerable due to the improved numerical robustness. However, variable, inhomogeneous filter widths feature significantly shorter computation times for solving the corresponding relaxation equation. Consequently, since inhomogeneous relaxation methods achieve better results with reduced computational effort, this approach will be further employed in the following. 

\subsection{Harbor Ferry -- High Froude-Number Case}
\label{subsec:hf}

\begin{figure}[!h]
\centering
\subfigure[]{
\iftoggle{tikzExternal}{
\input{./tikz/3D_harbor_ferry_scetch.tikz}
}{
\includegraphics{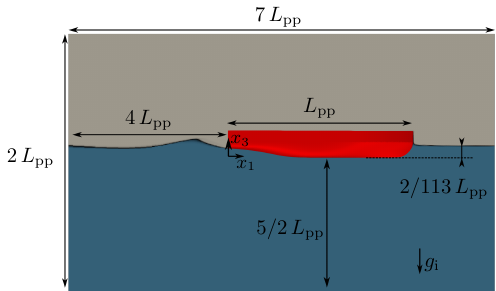}}
}
\subfigure[]{
\includegraphics[width=0.43\textwidth]{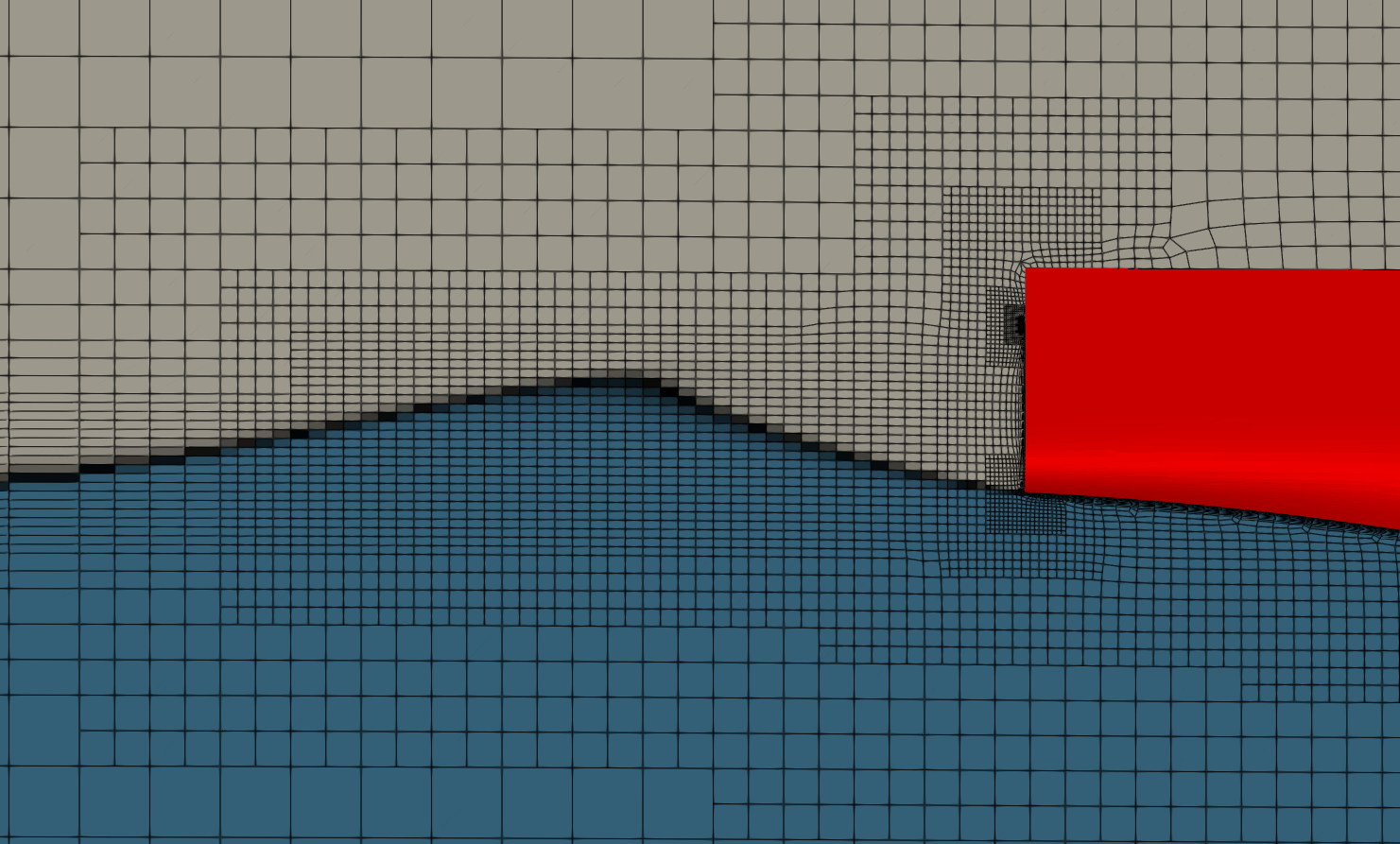}
}
\caption{Harbor Ferry Case case ($\mathrm{Re}_\mathrm{L} = \SI{2.43}{} \cdot 10^8$, $\mathrm{Fn}=0.4$): (a) Schematic drawing of the  initial configuration and (b) unstructured numerical grid around the  stern region.}
\label{fig:hf_scetch}
\end{figure}

Next, a test case with a considerably higher Froude number is considered. A Harbor Ferry cruising at $\mathrm{Fn}=0.4$ and $\mathrm{Re}_\mathrm{L} = \SI{2.43}{} \cdot 10^8$ based on the same similarity parameter setting as used for the JBC is investigated in full scale with active floatation. An exemplary sketch is provided in Fig. \ref{fig:hf_scetch}.  
The simulation domain has dimensions of $7 \, L_\mathrm{pp}$ in length, $2 \, L_\mathrm{pp}$ height, and $3 \, L_\mathrm{pp}$ in width, where outlet and bottom boundaries are positioned at distances of four as well as five-and-a-half hull lengths, respectively, from the origin. The expected dimensionless wavelength is $\lambda/ L_\mathrm{pp} = 2 \, \pi \, \mathrm{Fn}^2 = 1.0$, underlining the critical Froude-number setup.

\begin{figure}[!h]
\centering
\includegraphics[width=0.9\textwidth]{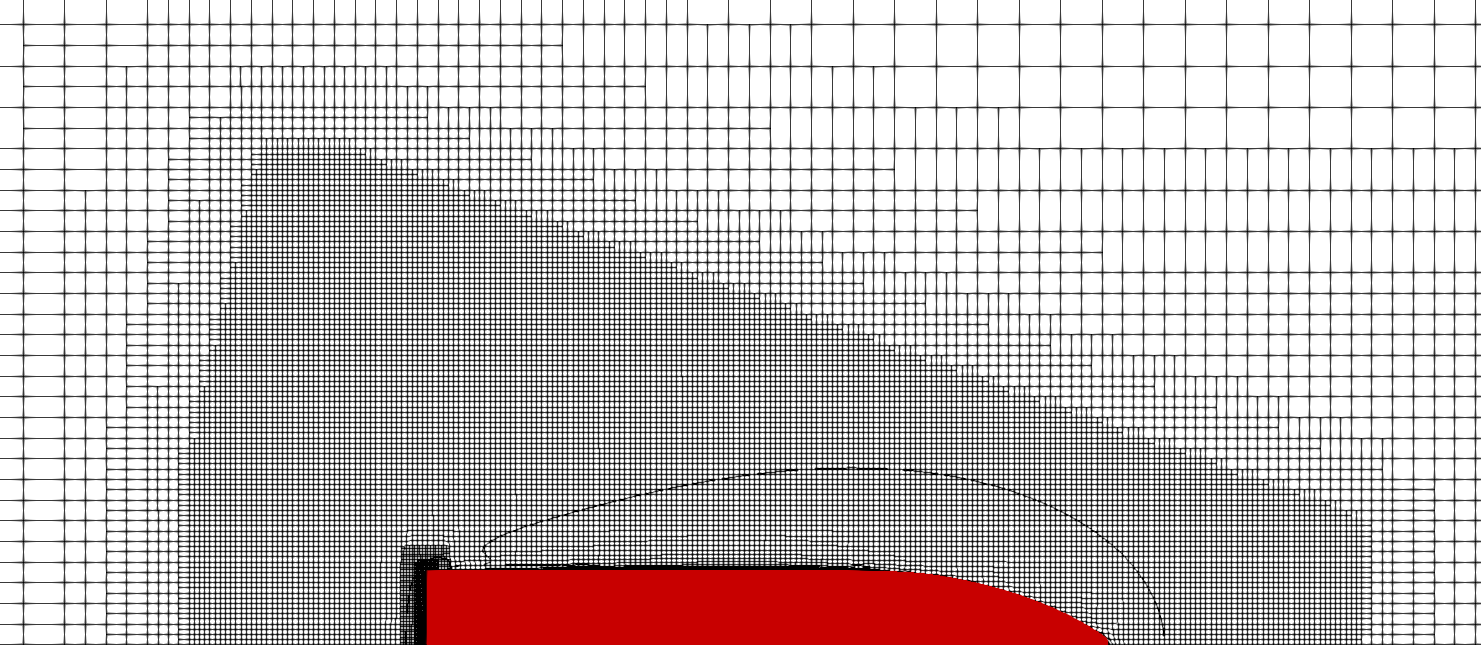}
\caption{Harbor Ferry Case case ($\mathrm{Re}_\mathrm{L} = \SI{2.43}{} \cdot 10^8$, $\mathrm{Fn}=0.4$):  Numerical grid in the still water plane.}
\label{fig:hf_free_surface}
\end{figure}

The discrete setup is similar to that of the JBC, especially the normalized surface (hull) and volume (free surface) refinements. Only the Kelvin wedge is less accurately captured, cf. Fig. \ref{fig:hf_free_surface}. The unstructured numerical grid comprises approximately $\SI{1.3}{} \cdot 10^6$ control volumes, cf. Fig. \ref{fig:hf_scetch} (b). Again, only half of the geometry is modeled in the lateral ($x_\mathrm{2}$) direction, and the simulations are advanced to a steady-state solution in pseudo-time with Courant numbers $\mathrm{Co} \leq 0.9$. The floatation is adjusted every \SI{2500}{} time steps as described in \cite{kuhl2021phd, kuhl2022adjoint}.

The fully consistent adjoint system could not be solved for the initial shape for both globally constant, homogeneous, and locally variable, inhomogeneous filter widths. This observation is typical for large-scale scenarios with high Reynolds and Froude numbers on unstructured numerical grids of an industrial process chain. All adjoint momentum sources in the non-wetted region are deactivated to circumvent this issue. This intervention usually stabilizes the adjoint calculations significantly since the extreme adjoint accelerations by the adjoint momentum sources co-occur in the free-surface-near region both in the air and the water but can be absorbed much better with less intense effects due to the density difference in the region of the latter. The approach is defensible, as the sensitivities in the area of the surface ship are minor, cf. Figs. \ref{fig:jbc_cinr_sens_diff}-\ref{fig:jbc_consistent_sens_diff} and \cite{kroger2016numerical, kroger2018adjoint}, and no convective transport occurs across the free-surface. The manipulation is implemented by simply scaling the adjoint momentum sources with the concentration field $c$ (or $1-c$ with an alternative volume fraction definition). It is important in this context that the manipulation is performed after the elliptical relaxation has been applied. Therefore, the redistribution mechanism of the relaxation procedure allows the transport of sources close to the free surface but in the air region across the air-water interface below the free water surface before deactivating less relevant but stability-reducing contributions above the free surface.

\begin{figure}[!htb]
\centering
\iftoggle{tikzExternal}{
\input{./tikz/3D_harbor_ferry_obj_type_1.tikz}}{
\includegraphics{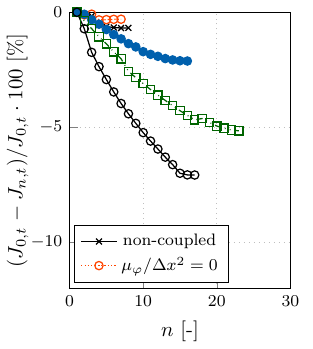}
\includegraphics{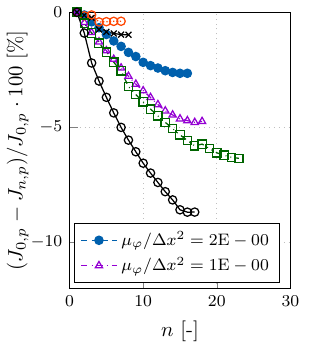}
\includegraphics{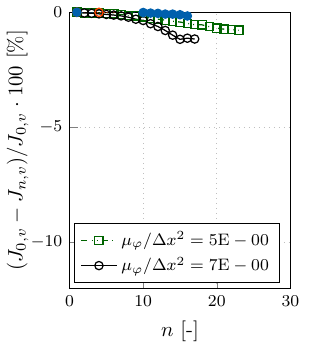}
}
\caption{Harbor Ferry Case case ($\mathrm{Re}_\mathrm{L} = \SI{2.43}{} \cdot 10^8$, $\mathrm{Fn}=0.4$):  Decrease in the objective functional over the number of optimization gradient steps for different elliptic relaxation strategies with varying filter homogeneous width of $\mu_\varphi / \Delta x^2 = [0, 1, 3, 5, 7]$.}
\label{fig:hf_obj_type_1}
\end{figure}

Based on the findings of the previous study at the JBC, the adjoint studies for the optimization of the harbor ferry are based exclusively on maximally consistent formulations (cf. experiment E4 from Sec. \ref{subsec:jbc}) and operate only with locally variable, inhomogeneous filter widths equal to or larger than the local grid resolution, i.e., $\mu_\varphi / \Delta x^2 = [0, 1, 3, 5, 7]$. Six shape optimizations are performed, considering the frozen ATC \& free surface optimization (cf. experiment E1 from Sec. \ref{subsec:jbc}) and the consistent one without relaxation. The optimizations again aim at minimizing the vessel's total resistance, and all optimizations choose their step size so that the first deformation field has a maximum displacement of $L_\mathrm{pp}/2000$. Figure \ref{fig:hf_obj_type_1} (left) plots the relative decrease in total resistance $(J_\mathrm{0,t} - J_\mathrm{n,t})/J_\mathrm{0,t} \cdot 100$ [\%] over the number of optimization steps. Additionally, the figure distinguishes between the pressure component $J_\mathrm{\cdot,p}$ acting normally to the hull (center) and the tangentially acting friction component $J_\mathrm{\cdot,v}$ (right).
It can be seen that the decrease in the cost functional increases for larger filter widths. This initially unexpected observation follows from the source term manipulation based on the concentration field and the interaction with the elliptical relaxation method. Since a substantial part of the adjoint source terms originates from the free-surface-near region and is primarily canceled at low relaxation length due to the manipulation in the air region, significantly more adjoint momentum source contributions migrate into the completely wetted region for larger filter widths and thus advantageously enrich the adjoint velocity field under water -- which enters the sensitivity expression.
The adjoint system without equation cross-coupling only achieves a resistance reduction of approx. 1\% that seems comparatively low compared to previously published results for similar ships at high Froude numbers and full-scale Reynolds numbers, cf. \cite{kuhl2022adjoint}. However, the consistent optimization without elliptic relaxation diverges quickly and achieves even only half the gain of the non-consistent optimization. All other consistent optimizations with a non-zero filter width lead to significantly higher drag reductions of about 2.5\% up to 7.5\% for $\mu_\varphi / \Delta x^2 = 1$ and $\mu_\varphi / \Delta x^2 = 7$, respectively. The optimization procedure primarily changes the pressure component of the total resistance.

\begin{figure}[!htb]
\centering
\iftoggle{tikzExternal}{
\input{./tikz/3D_harbor_ferry_hydrostatics.tikz}}{
\includegraphics{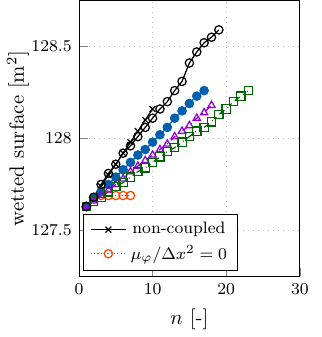}
\includegraphics{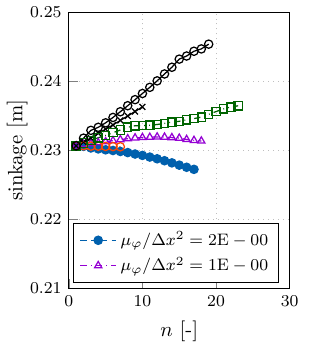}
\includegraphics{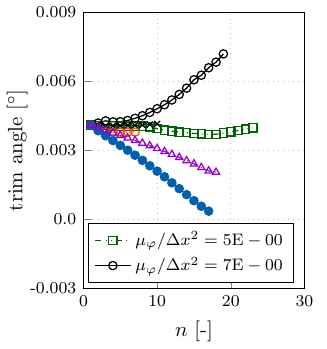}
}
\caption{Harbor Ferry Case case ($\mathrm{Re}_\mathrm{L} = \SI{2.43}{} \cdot 10^8$, $\mathrm{Fn}=0.4$): Wetted surface (left), sinkage (center), and trim angle (right) over the number of optimization gradient steps for different elliptic relaxation strategies with varying filter homogeneous width of $\mu_\varphi / \Delta x^2 = [0, 1, 3, 5, 7]$.}
\label{fig:hf_hydrostatics}
\end{figure}

Figure \ref{fig:hf_hydrostatics} shows the development of typical hydrostatic variables over the optimization step counter, i.e., the wetted area (left), the sinkage (center), and the trim angle (right). A clear trend can be observed only in the wetted surface, which increases in all cases. However, both the sinkage and the trim behave differently in the trend, depending on which relaxation method is used, with small [large] filter widths leading to a reduction [increase] in the sinkage and trim.
In particular, the latter two aspects of the vessel's floatation are sensitive to the small disturbances in the Froude number, i.e., the hull length in this optimization study, which was deliberately not fixed.

\begin{figure}[!htb]
\centering
\iftoggle{tikzExternal}{
\input{./tikz/3D_harbor_ferry_slices.tikz}}{
\includegraphics{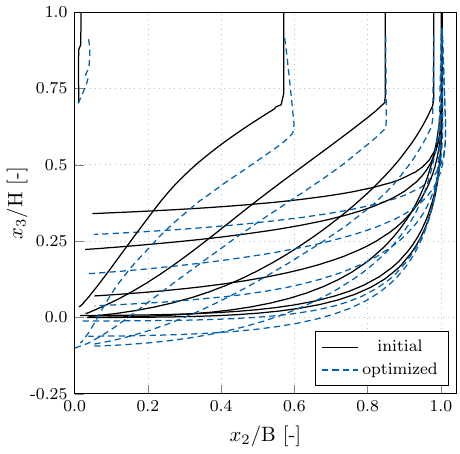}
\includegraphics{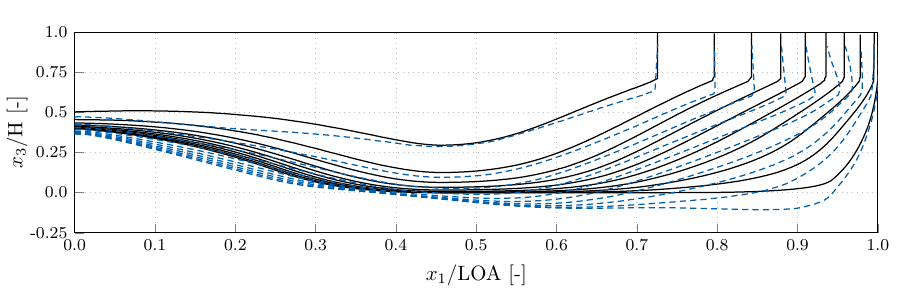}
\includegraphics{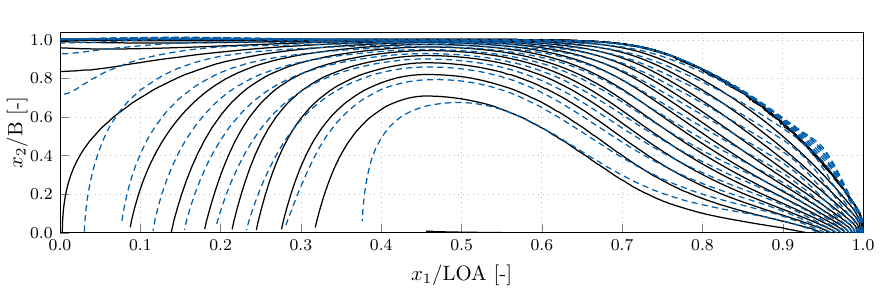}
}
\caption{Harbor Ferry Case case ($\mathrm{Re}_\mathrm{L} = \SI{2.43}{} \cdot 10^8$, $\mathrm{Fn}=0.4$): Frames (top), water lines (center), and buttocks (bottom) of the initial (black solid) harbor ferry and the with a filter width of $\mu_\varphi / \Delta x^2 = 7$ optimized version (blue dashed).}
\label{fig:hf_slices}
\end{figure}

The optimized harbor ferry based on a filter width of $\mu_\varphi / \Delta x^2 = 7$ is compared against the initial shape in Fig. \ref{fig:hf_slices} based on the frames (top), water lines (center), and buttocks (bottom), where LOA, B, and H refer to the initial vessel's length overall, width and height, respectively. The increase in draught and minimization of the S-shape at the transition from the midship to the stern becomes apparent in the waterlines. The buttocks also indicate an elongation of the bow area.

For verification purposes, the final shape was meshed again with the setup of the initial geometry (cf. the beginning of the section) and simulated from scratch. The results are almost identical to those of the optimization study.

\section{Conclusion}
\label{sec:conclusion}

The article investigated a method for the numerical stabilization of sequential continuous adjoint fluid dynamic simulation methods based on elliptic relaxation strategies. The approach is based on a Partial Differential Equation (PDE) containing a single parameter to be specified by the user. Analytical investigations revealed that this parameter can be interpreted as the filter width of a probabilistic density function or a Gaussian kernel. Important properties of the approach are associated with (a) smoothing effects that feature redistribution capabilities with (b) simultaneous integral impact conservation. Both aspects have been numerically validated and applied to those numerically unpleasant adjoint equation cross-coupling terms that trigger numerical instabilities due to their explicit consideration in the underlying sequential solution procedure.

Generally, the smoothing parameter can be chosen arbitrarily, whereby the studies in the paper proposed a coupling to (a) inhomogeneous, local spatial discretization measures or (b) homogeneous, global reference lengths. Numerical stabilization effects were achieved in both cases, especially concerning the often-discussed Adjoint Transpose Convection (ATC) term. However, a coupling to the local grid resolution provided an improved behavior and turned out to be the more economical approach. When scaling the filter width with the local grid spacing, the order of magnitude should always be as large as necessary and as small as possible so as not to compromise the adjoint-based sensitivity's consistency. Filter widths that were a single-digit multiple of the local grid resolution resulted in mean relative errors in the single-digit percentage range for the test and application cases investigated in the paper, which was sufficient to stabilize the adjoint process in such a way that higher cost functional reductions were achieved compared adjoint strategies that feature improved stability by neglecting the numerically unpleasant explicit equation cross-coupling terms. Consequently, regarding numerical stability, this inhomogeneous filter strategy outperformed all other approaches and can be recommended as best practice.

In the case of two-phase full-scale ship flow studies at high Froude and  Reynolds numbers, a substantial optimization could only be achieved by a solid manipulation of the adjoint source terms in the non-wetted domain combined with the relaxation method presented.

\section{Declaration of Competing Interest}
The author declares that he has no known competing financial interests or personal relationships that could have appeared to influence the work reported in this paper.

\section{Acknowledgments}
The current work is part of the "Propulsion Optimization of Ships and Appendages" research project, which is funded by the German Federal Ministry for Economics and Climate Action (Grant No. 03SX599C). The author gratefully acknowledges this support.
The author would also like to thank the engineers at J.M. Voith SE \& Co. KG for providing the challenging optimization case of the harbor ferry.
The capabilities of Wolfram Research Inc.'s Mathematica Online tool (\cite{mathematicaOnline}) for the analytical solution of ordinary differential equations are acknowledged.


\begin{thebibliography}{45}
\providecommand{\natexlab}[1]{#1}
\providecommand{\url}[1]{\texttt{#1}}
\expandafter\ifx\csname urlstyle\endcsname\relax
  \providecommand{\doi}[1]{doi: #1}\else
  \providecommand{\doi}{doi: \begingroup \urlstyle{rm}\Url}\fi

\bibitem[Al-Jamal et~al.(2018)Al-Jamal, Alomari, and
  Gockenbach]{al2018smoothing}
M.F. Al-Jamal, A.K. Alomari, and M.S. Gockenbach.
\newblock {Smoothing via Elliptic Operators with Application to Edge
  Detection}.
\newblock \emph{Inverse Problems in Science and Engineering}, 26\penalty0
  (5):\penalty0 657--676, 2018.
\newblock \doi{0.1080/17415977.2017.1336552}.

\bibitem[Beckers et~al.(2019)Beckers, Behrens, and Wollner]{beckers2019duality}
S.~Beckers, J.~Behrens, and W.~Wollner.
\newblock Duality {B}ased {E}rror {E}stimation in the {P}resence of
  {D}iscontinuities.
\newblock \emph{Applied Numerical Mathematics}, 144:\penalty0 83--99, 2019.
\newblock \doi{10.1016/j.apnum.2019.05.016}.

\bibitem[Borrvall and Petersson(2003)]{borrvall2003topology}
T.~Borrvall and J.~Petersson.
\newblock {Topology Optimization of Fluids in Stokes Flow}.
\newblock \emph{International Journal for Numerical Methods in Fluids},
  41\penalty0 (1):\penalty0 77--107, 2003.

\bibitem[Dick et~al.(2022)Dick, Gauger, and Schmidt]{dick2022combining}
T.~Dick, N.R. Gauger, and S~Schmidt.
\newblock Combining {S}obolev {S}moothing with {P}arameterized {S}hape
  {O}ptimization.
\newblock \emph{Computers \& Fluids}, 244:\penalty0 105568, 2022.
\newblock ISSN 0045-7930.
\newblock \doi{https://doi.org/10.1016/j.compfluid.2022.105568}.

\bibitem[Durbin(1991)]{durbin1991near}
P.A. Durbin.
\newblock Near-{W}all {T}urbulence {C}losure {M}odeling {W}ithout “{D}amping
  {F}unctions”.
\newblock \emph{Theoretical and Computational Fluid Dynamics}, 3\penalty0
  (1):\penalty0 1--13, 1991.

\bibitem[Galanos et~al.(2022)Galanos, Papoutsis-Kiachagias, Giannakoglou,
  Kondo, and Tanimoto]{galanos2022synergistic}
N.~Galanos, E.M. Papoutsis-Kiachagias, K.C. Giannakoglou, Y.~Kondo, and
  K.~Tanimoto.
\newblock {Synergistic Use of Adjoint-Based Topology and Shape Optimization for
  the Design of Bi-Fluid Heat Exchangers}.
\newblock \emph{Structural and Multidisciplinary Optimization}, 65\penalty0
  (9):\penalty0 245, 2022.
\newblock \doi{10.1007/s00158-022-03330-w}.

\bibitem[Giles and Pierce(2000)]{giles2000introduction}
M.B. Giles and N.A. Pierce.
\newblock An {I}ntroduction to the {A}djoint {A}pproach to {D}esign.
\newblock \emph{Flow, Turbulence and Combustion}, 65\penalty0 (3):\penalty0
  393--415, 2000.
\newblock \doi{10.1023/A:1011430410075}.

\bibitem[Grossmann(1997)]{grossmann1997smoothing}
C~Grossmann.
\newblock {Smoothing of Elliptic Differential Inclusions and its Iterative
  Treatment}.
\newblock \emph{Numerical Functional Analysis and Optimization}, 18\penalty0
  (1-2):\penalty0 93--105, 1997.
\newblock \doi{0.1080/01630569708816749}.

\bibitem[Haubner et~al.(2020)Haubner, Ulbrich, and
  Ulbrich]{haubner2020analysis}
J.~Haubner, M.~Ulbrich, and S.~Ulbrich.
\newblock {Analysis of Shape Optimization Problems for Unsteady Fluid-Structure
  Interaction}.
\newblock \emph{Inverse Problems}, 36:\penalty0 1--38, 2020.

\bibitem[Hino et~al.(2020)Hino, Stern, Larsson, Visonneau, Hirata, and
  Kim]{hino2020numerical}
T.~Hino, F.~Stern, L.~Larsson, M.~Visonneau, N.~Hirata, and J.~Kim.
\newblock \emph{{Numerical Ship Hydrodynamics: An Assessment of the Tokyo 2015
  Workshop}}, volume~94.
\newblock Springer Nature, 2020.

\bibitem[Hojjat et~al.(2014)Hojjat, Stavropoulou, and
  Bletzinger]{hojjat2014vertex}
M.~Hojjat, E.~Stavropoulou, and K.-U. Bletzinger.
\newblock The {V}ertex {M}orphing method for node-based shape optimization.
\newblock \emph{Computer Methods in Applied Mechanics and Engineering},
  268:\penalty0 494--513, 2014.

\bibitem[Jameson(2003)]{jameson2003aerodynamic}
A.~Jameson.
\newblock {Aerodynamic Shape Optimization Using the Adjoint Method}, 2003.

\bibitem[Karpouzas et~al.(2016)Karpouzas, Papoutsis-Kiachagias, Schumacher,
  Villiers, Giannakoglou, and C.]{karpouzas2016adjoint}
G.K. Karpouzas, E.~M. Papoutsis-Kiachagias, T.~Schumacher, E.~Villiers, K.~C.
  Giannakoglou, and Othmer C.
\newblock Adjoint {O}ptimization for {V}ehicle {E}xternal {A}erodynamics.
\newblock \emph{International Journal of Automotive Engineering}, 7\penalty0
  (1):\penalty0 1--7, 2016.

\bibitem[Kim et~al.(2005)Kim, Hosseini, Jameson, and
  Leoviriyakit]{kim2005enhancement}
S.~Kim, K.~Hosseini, A.~Jameson, and K.~Leoviriyakit.
\newblock {Enhancement of Adjoint Design Methods via Optimization of Adjoint
  Parameters}.
\newblock In \emph{43rd AIAA Aerospace Sciences Meeting and Exhibit}, page 448,
  2005.

\bibitem[Kr\"oger(2016)]{kroger2016numerical}
J.~Kr\"oger.
\newblock \emph{A {N}umerical {P}rocess for the {H}ydrodynamic {O}ptimisation
  of {S}hips}.
\newblock PhD thesis, Hamburg University of Technology, 2016.

\bibitem[Kr{\"o}ger and Rung(2016)]{kroger2015cad}
J.~Kr{\"o}ger and T.~Rung.
\newblock {CAD}-{F}ree {H}ydrodynamic {O}ptimisation {U}sing {C}onsistent
  {K}ernel-{B}ased {S}ensitivity {F}iltering.
\newblock \emph{Ship Technology Research}, 62\penalty0 (3):\penalty0 111--130,
  2016.
\newblock \doi{10.1080/09377255.2015.1109872}.

\bibitem[Kr{\"o}ger et~al.(2018)Kr{\"o}ger, K{\"u}hl, and
  Rung]{kroger2018adjoint}
J.~Kr{\"o}ger, N.~K{\"u}hl, and T.~Rung.
\newblock Adjoint {V}olume-of-{F}luid {A}pproaches for the {H}ydrodynamic
  {O}ptimisation of {S}hips.
\newblock \emph{Ship Technology Research}, 65\penalty0 (1):\penalty0 47--68,
  January 2018.
\newblock \doi{10.1080/09377255.2017.1411001}.

\bibitem[K{\"u}hl(2021)]{kuhl2021phd}
N.~K{\"u}hl.
\newblock \emph{Adjoint-{B}ased {S}hape {O}ptimization {C}onstraint by
  {T}urbulent {T}wo-{P}hase {N}avier-{S}tokes {S}ystems}.
\newblock PhD thesis, Hamburg University of Technology, 2021.

\bibitem[K{\"u}hl and Rung(2022)]{kuhl2022discrete}
N.~K{\"u}hl and T.~Rung.
\newblock Discrete {A}djoint {M}omentum-{W}eighted {I}nterpolation
  {S}trategies.
\newblock \emph{Journal of Computational Physics}, 467:\penalty0 111474, 2022.
\newblock \doi{10.1016/j.jcp.2022.111474}.

\bibitem[K{\"u}hl et~al.(2021{\natexlab{a}})K{\"u}hl, Kr{\"o}ger, Siebenborn,
  Hinze, and Rung]{kuhl2021adjoint}
N.~K{\"u}hl, J.~Kr{\"o}ger, M.~Siebenborn, M.~Hinze, and T.~Rung.
\newblock Adjoint {C}omplement to the {V}olume-of-{F}luid {M}ethod for
  {I}mmiscible {F}lows.
\newblock \emph{Journal of Computational Physics}, 440:\penalty0 110411,
  2021{\natexlab{a}}.
\newblock \doi{10.1016/j.jcp.2021.110411}.

\bibitem[K{\"u}hl et~al.(2021{\natexlab{b}})K{\"u}hl, M{\"u}ller, and
  Rung]{kuhl2021continuous}
N.~K{\"u}hl, P.M. M{\"u}ller, and T.~Rung.
\newblock Continuous {A}djoint {C}omplement to the {B}lasius {E}quation.
\newblock \emph{Physics of Fluids}, 33\penalty0 (3):\penalty0 033608,
  2021{\natexlab{b}}.
\newblock \doi{10.1063/5.0037779}.

\bibitem[K{\"u}hl et~al.(2022)K{\"u}hl, Nguyen, Palm, J{\"u}rgens, and
  Rung]{kuhl2022adjoint}
N.~K{\"u}hl, T.~T. Nguyen, M.~Palm, D.~J{\"u}rgens, and T.~Rung.
\newblock Adjoint {N}ode-{B}ased {S}hape {O}ptimization of {F}ree {F}loating
  {V}essels.
\newblock \emph{Structural and Multidisciplinary Optimization}, 65:\penalty0
  247, 2022.
\newblock \doi{10.1007/s00158-022-03338-2}.

\bibitem[Lazarov and Sigmund(2011)]{lazarov2011filters}
B.S. Lazarov and O.~Sigmund.
\newblock {Filters in Topology Optimization Based on Helmholtz-Type
  Differential Equations}.
\newblock \emph{International Journal for Numerical Methods in Engineering},
  86\penalty0 (6):\penalty0 765--781, 2011.
\newblock \doi{10.1002/nme.3072}.

\bibitem[Lazarov et~al.(2016)Lazarov, Wang, and Sigmund]{lazarov2016length}
B.S. Lazarov, F.~Wang, and O.~Sigmund.
\newblock {Length Scale and Manufacturability in Density-Based Topology
  Optimization}.
\newblock \emph{Archive of Applied Mechanics}, 86:\penalty0 189--218, 2016.
\newblock \doi{10.1007/s00419-015-1106-4}.

\bibitem[L{\"o}hner et~al.(2003)L{\"o}hner, Soto, and Yang]{lohner2003adjoint}
R.~L{\"o}hner, O.~Soto, and C.~Yang.
\newblock An {A}djoint-{B}ased {D}esign {M}ethodology for {CFD} {O}ptimization
  {P}roblems.
\newblock In \emph{41st Aerospace Sciences Meeting and Exhibit, Reno, Nevada},
  page 299, 2003.

\bibitem[Manzke(2018)]{manzke2018development}
M.~Manzke.
\newblock \emph{Development of a {S}calable {M}ethod for the {E}fficient
  {S}imulation of {F}lows using {Dynamic} {G}oal-{O}rientated {Local}
  {Grid}-{Adaption}}.
\newblock PhD thesis, Hamburg University of Technology, 2018.

\bibitem[Margetis et~al.(2021)Margetis, Papoutsis-Kiachagias, and
  Giannakoglou]{margetis2021lossy}
A.-S.I. Margetis, E.M. Papoutsis-Kiachagias, and K.C. Giannakoglou.
\newblock Lossy {C}ompression {T}echniques {S}upporting {U}nsteady {A}djoint on
  2d/3d {U}nstructured {G}rids.
\newblock \emph{Computer Methods in Applied Mechanics and Engineering},
  387:\penalty0 114152, 2021.
\newblock \doi{10.1016/j.cma.2021.114152}.

\bibitem[Marta and Shankaran(2013)]{marta2013handling}
A.C. Marta and S.~Shankaran.
\newblock On the {H}andling of {T}urbulence {E}quations in {RANS} {A}djoint
  {S}olvers.
\newblock \emph{Computers \& Fluids}, 74:\penalty0 102--113, 2013.
\newblock \doi{10.1016/j.compfluid.2013.01.012}.

\bibitem[Menter(1994)]{menter1994two}
F.R. Menter.
\newblock Two-{E}quation {E}ddy-{V}iscosity {T}urbulence {M}odels for
  {E}ngineering {A}pplications.
\newblock \emph{AIAA Journal}, 32\penalty0 (8):\penalty0 1598--1605, 1994.
\newblock \doi{10.2514/3.12149}.

\bibitem[Mohammadi and Pironneau(2010)]{mohammadi2010applied}
B.~Mohammadi and O.~Pironneau.
\newblock \emph{Applied {S}hape {O}ptimization for {F}luids}.
\newblock Oxford University Press, 2010.

\bibitem[Nadarajah(2003)]{nadarajah2003discrete}
S.~K. Nadarajah.
\newblock \emph{The {D}iscrete {A}djoint {A}pproach to {A}erodynamic {S}hape
  {O}ptimization}.
\newblock PhD thesis, Stanford University, 2003.

\bibitem[Nadarajah and Jameson(2000)]{nadarajah2000comparison}
S.~K. Nadarajah and A.~Jameson.
\newblock A {C}omparison of the {C}ontinuous and {D}iscrete {A}djoint
  {A}pproach to {A}utomatic {A}erodynamic {O}ptimization.
\newblock In \emph{38th Aerospace Sciences Meeting and Exhibit, Reno, Nevada},
  AIAA--2000--0667, 2000.
\newblock \doi{10.2514/6.2000-667}.

\bibitem[Nielsen et~al.(2004)Nielsen, Lu, Park, and
  Darmofal]{nielsen2004implicit}
E.J. Nielsen, J.~Lu, M.A. Park, and D.L. Darmofal.
\newblock An {I}mplicit, {E}xact {D}ual {A}djoint {S}olution {M}ethod for
  {T}urbulent {F}lows on {U}nstructured {G}rids.
\newblock \emph{Computers \& Fluids}, 33\penalty0 (9):\penalty0 1131--1155,
  2004.
\newblock \doi{10.1016/j.compfluid.2003.09.005}.

\bibitem[Othmer(2008)]{othmer2008continuous}
C.~Othmer.
\newblock A {C}ontinuous {A}djoint {F}ormulation for the {C}omputation of
  {T}opological and {S}urface {S}ensitivities of {D}ucted {F}lows.
\newblock \emph{International Journal for Numerical Methods in Fluids},
  58\penalty0 (8):\penalty0 861--877, 2008.
\newblock \doi{10.1002/fld.1770}.

\bibitem[Othmer(2014)]{othmer2014adjoint}
C.~Othmer.
\newblock Adjoint {M}ethods for {C}ar {A}erodynamics.
\newblock \emph{Journal of Mathematics in Industry}, 4\penalty0 (1):\penalty0
  6, 2014.
\newblock \doi{10.1186/2190-5983-4-6}.

\bibitem[Peter and Dwight(2010)]{peter2010numerical}
J.E.V. Peter and R.P. Dwight.
\newblock Numerical {S}ensitivity {A}nalysis for {A}erodynamic {O}ptimization:
  {A} {S}urvey of {A}pproaches.
\newblock \emph{Computers \& Fluids}, 39\penalty0 (3):\penalty0 373--391, 2010.
\newblock \doi{10.1016/j.compfluid.2009.09.013}.

\bibitem[Pope(2001)]{pope2001turbulent}
S.B. Pope.
\newblock \emph{Turbulent {F}lows}.
\newblock Cambridge University Press, 2001.

\bibitem[Roth and Ulbrich(2013)]{roth2013discrete}
R.~Roth and S.~Ulbrich.
\newblock {A Discrete Adjoint Approach for the Optimization of Unsteady
  Turbulent Flows}.
\newblock \emph{Flow, Turbulence and Combustion}, 90:\penalty0 763--783, 2013.
\newblock \doi{10.1007/s10494-012-9439-3}.

\bibitem[Rung et~al.(2009)Rung, W{\"o}ckner, Manzke, Brunswig, Ulrich, and
  St{\"u}ck]{rung2009challenges}
T.~Rung, K.~W{\"o}ckner, M.~Manzke, J.~Brunswig, C.~Ulrich, and A.~St{\"u}ck.
\newblock Challenges and {P}erspectives for {M}aritime {CFD} {A}pplications.
\newblock \emph{Jahrbuch der Schiffbautechnischen Gesellschaft}, 103:\penalty0
  127--39, 2009.

\bibitem[Schubert(2019)]{schubert2019analysis}
S.~Schubert.
\newblock \emph{Analysis of {C}oupling {T}echniques for {O}verset-{G}rid
  {F}inite-{V}olume {M}ethods}.
\newblock PhD thesis, Hamburg University of Technology, 2019.

\bibitem[St{\"u}ck(2012)]{stuck2012adjoint}
A.~St{\"u}ck.
\newblock \emph{Adjoint {N}avier-{S}tokes {M}ethods for {H}ydrodynamic {S}hape
  {O}ptimisation}.
\newblock PhD thesis, Hamburg University of Technology, 2012.

\bibitem[St{\"u}ck and Rung(2011)]{stuck2011adjoint}
A.~St{\"u}ck and T.~Rung.
\newblock Adjoint {RANS} with {F}iltered {S}hape {D}erivatives for
  {H}ydrodynamic {O}ptimisation.
\newblock \emph{Computers \& Fluids}, 47\penalty0 (1):\penalty0 22--32, 2011.
\newblock \doi{10.1016/j.compfluid.2011.01.041}.

\bibitem[St{\"u}ck and Rung(2013)]{stuck2013adjoint}
A.~St{\"u}ck and T.~Rung.
\newblock Adjoint {C}omplement to {V}iscous {F}inite-{V}olume
  {P}ressure-{C}orrection {M}ethods.
\newblock \emph{Journal of Computational Physics}, 248:\penalty0 402--419,
  2013.
\newblock \doi{10.1016/j.jcp.2013.01.002}.

\bibitem[Wilcox(1998)]{wilcox1998turbulence}
D.C. Wilcox.
\newblock \emph{Turbulence {M}odeling for {CFD}}, volume~2.
\newblock DCW Industries La Canada, 1998.

\bibitem[{Wolfram Research Inc.}(2024)]{mathematicaOnline}
{Wolfram Research Inc.}
\newblock Mathematica online, {V}ersion 14.1, 2024.
\newblock URL \url{https://www.wolfram.com/mathematica}.
\newblock Champaign, IL.

\end{thebibliography}
\end{document}